%% file: main.tex
\documentclass{lmcs}

\input{packages}
\input{macros}

\begin{document}


  \title[sscc]{A Rewriting Logic Approach to Stochastic and Spatial
    Constraint System Specification and Verification}
  

  \author[Romero M.]{Miguel Romero}
  \address{Departament of Electronics and Computer Science Pontificia Universidad Javeriana Cali, Colombia}

  \author[Ram\'irez S.]{Sergio Ram\'{i}rez}
  \address{Universidad EAFIT, Medellín, Colombia}

  \author[Rocha C.]{Camilo Rocha}
  \address{Departament of Electronics and Computer Science Pontificia Universidad Javeriana Cali, Colombia}

  \author[Valencia F.]{Frank Valencia}
  \address{Departament of Electronics and Computer Science Pontificia Universidad Javeriana Cali, Colombi
  CNRS, LIX \'Ecole Polytechnique de Paris, France}



  \begin{abstract}
    This paper addresses the issue of specifying, simulating, and
    verifying reactive systems in rewriting logic. It presents an
    executable semantics for probabilistic, timed, and spatial
    concurrent constraint programming ---here called \textit{stochastic
      and spatial concurrent constraint} systems ($\SSCC$)--- in the
    rewriting logic semantic framework. The approach is based on an
    enhanced and generalized model of \textit{concurrent constraint
      programming} ($\CCP$) where computational hierarchical spaces can
    be assigned to belong to agents. The executable semantics faithfully
    represents and operationally captures the highly concurrent nature,
    uncertain behavior, and spatial and epistemic characteristics of
    reactive systems with flow of information. In $\SSCC$, timing
    attributes ---represented by stochastic duration--- can be
    associated to processes, and exclusive and independent probabilistic
    choice is also supported. SMT solving technology, available from the
    Maude system, is used to realize the underlying constraint system of
    $\SSCC$ with quantifier-free formulas over integers and reals.  This
    results in a fully executable real-time symbolic specification that
    can be used for quantitative analysis in the form of statistical
    model checking. The main features and capabilities of $\SSCC$ are
    illustrated with examples throughout the paper. This contribution is
    part of a larger research effort aimed at making available formal
    analysis techniques and tools, mathematically founded on the $\CCP$
    approach, to the research community.
  \end{abstract}



\maketitle

\section{Introduction}
\label{sec.intro}

Reactive systems are a broad class of concurrent systems used to meet
today's application demands in, e.g., cloud-based clusters with
hundreds of processors, 100\% uptime systems with milliseconds time
response, and petabytes of information produced by social networks in
very short periods of time. According to the Reactive
Manifesto~\cite{boner-reacmanif-2014}, a reactive system is a
responsive, resilient, elastic, and message driven system that is
almost the standard of many real-world applications these
days. Responsive systems are about establishing reliable upper bounds
so that consistent behavior builds user confidence because of rapid
response times to external stimuli, also requiring that such stimuli
from the unpredictable environment is dealt with properly. Resilience
refers to a property that all mission-critical systems must meet:
responsiveness to failure. This means that even when a failure
happens, the system must continue to be responsive and meet the user
requirements. Elasticity is a key feature of reactive systems that,
under varying workload, can increase or decrease allocated resources.
Message driven means that component interaction within a system relies
on asynchronous message passing allowing, for instance, non-blocking
communication in which components only consume resources while
active.

The importance and proliferation of reactive systems in today's world
is central to this research. From an abstract standpoint, the reason
why reactive systems are of special interest is that they can
illustrate the complexity and nature of today's ubiquitous computing
where many agents execute and share information in a distributed
configuration. Although this characterization is widely generic, it
gives key insights on the importance and inherent complexity of
reactive systems and justifies why they are gaining ground, both as an
industrial paradigm and as a subject of research. However, giving
mathematical foundations and formal meaning to reactive computation is
a serious challenge since traditional mathematical models of
computation do not single out important aspects of these systems such
as information flow and hierarchical agent structures. Furthermore,
without the required mathematical scaffolding, quantitative analysis
useful to understand or predict, e.g., the behavior, reliability, and
responsiveness ---which are key attributes of correctness and
utility--- is well beyond the reach of any serious practitioner or
industry. Therefore, the question of how to specify, simulate, and
verify a reactive system is an important one, especially in the
presence of quantitative demands.

Key features of reactive systems have been sufficiently addressed in
the context of concurrent constraint programming
($\CCP$)~\cite{saraswat-ccpsem-1991}, a well-established process model
for concurrency based upon the shared-variables communication model.
Its basic intuitions arise mostly from logic; in fact, $\CCP$
processes can be interpreted both as concurrent computational entities
and logic specifications (e.g., process composition can be seen as
parallel execution and as conjunction). In $\CCP$, agents can interact
by \emph{posting} (or \emph{telling}) partial information in a medium
such as a centralized \emph{store}. Partial information is represented
by constraints on the shared variables of the system. The other way in
which agents can interact is by \emph{querying} (or \emph{asking})
about partial information entailed by the store. This provides the
synchronization mechanism of the model: asking agents are suspended
until there is enough information in the store to answer their
query. As other mature models of concurrency, $\CCP$ has been extended
to capture aspects such as
mobility~\cite{gilbert-ccpmob-2000,guzman-sccpe-2017}, and ---most
prominently---
probabilistic~\cite{gupta-pcc-1997,gupta-stochastic-1999,perez-pntcc-2008,aranda-pntccsos-2008}
and
temporal~\cite{saraswat-tcc-1994,saraswat-deftcc-1995,deboer-timedcc-2000,nielsen-temporal-2002,perez-pntcc-2008,chiarugi-stochmodel-2013}
reactive computation, where processes can be constrained also by
probabilistic choice, unit delays, and time-out conditions.

Due to their centralized notion of store, all the previously-mentioned
extensions are unsuitable for today's systems where information and
processes can be spatially distributed among certain groups of
agents. Examples of these systems include agents posting and querying
information in the presence of \emph{spatial hierarchies for sharing
 information and knowledge}, such as friend circles and shared albums
in social networks, or shared folders in cloud storage. Recently, the
authors of~\cite{knight-sccp-2012} enhanced and generalized the theory
of $\CCP$ to systems with spatial distribution of information in the
novel \emph{spatial constraint system} ($\SCS$), where computational
hierarchical spaces can be assigned to belong to agents. In $\SCS$,
each space may have $\CCP$ processes and other sub-spaces, and
processes can post and query information in their given space (i.e.,
locally) and may as well move from one space to another.

This work is part of a larger research effort aimed at making
available formal analysis techniques and tools, mathematically founded
on the $\CCP$ approach, to the reactive systems community. The goal of
the present paper is to introduce an executable semantics for
probabilistic, timed, and spatial concurrent constraint systems, here
called \textit{stochastic and spatial concurrent constraint} systems
($\SSCC$). Towards this endeavor, rewriting
logic~\cite{meseguer-rltcs-1992} is used as a semantic framework to
faithfully represent and operationally capture the highly concurrent
nature, uncertain behavior, and spatial and epistemic spirit of
reactive systems in a robust and executable specification. This
specification, by being executable in the Maude
system~\cite{clavel-maudebook-2007}, is amenable to automatic
algorithmic quantitative analysis in the form of statistical model
checking.

The $\SSCC$ executable rewriting logic semantics unifies and extends
previous efforts by the research community to capture phenomena of
reactive systems in the $\CCP$ mathematical framework. The notions of
timed behavior and probabilistic choice, and non-determinism in
$\textsf{tcc}$~\cite{saraswat-tcc-1994} and
$\textsf{pntcc}$~\cite{perez-pntcc-2008}, respectively, can be well
expressed in $\SSCC$. The notion of processes with stochastic
duration, as proposed in~\cite{aranda-pntccsos-2008}, is also
expressible in the proposed semantics. In $\SSCC$, the underlying
notion of hierarchical and distributed store found in
\textsf{scc}~\cite{knight-sccp-2012} is faithfully modeled by local
spaces, where inconsistent local stores need not propagate their
inconsistencies towards the global store. Previous efforts by the
authors to bring executable semantics and reachability analysis to
$\textsf{scc}$ using rewriting logic in~\cite{romero-sccpnfm-2018}
---and its subsequent extension to real-time
in~\cite{ramirez-rtsccpwrla-2018}--- are transparently subsumed as
sub-models in $\SSCC$, and further extended with stochastic features
and support for statistical model checking. This latter claim means
that the $\SSCC$ executable semantics presented in this work can also
be used for reachability analysis and LTL model checking of reactive
systems, although these features are not part of the present
exposition. The reader is referred
to~\cite{romero-sccpnfm-2018,ramirez-rtsccpwrla-2018} for an
explanation and examples of how these features can be supported by a
subset of the rewriting logic semantics presented in this work.

In the $\SSCC$ rewriting logic semantics, flat configurations of
object-like terms encode the hierarchical structure of spaces, and
equational and rewrite rules axiomatize the concurrent computational
steps of processes. Time attributes are associated to process-store
interaction with stochastic duration, as well as to process mobility
in the space structure, by means of maps from agents to probability
distribution functions. These choices can be interpreted to denote, as
previously mentioned, upper bounds in the execution time of the given
operations. Furthermore, exclusive and independent non-determinism,
which complement the existing form of non-deterministic choice in
$\CCP$, can be parametric on probabilistic choice, modeling the fact
that processes can sometimes execute due to interaction with the
external environment. The underlying constraint system of $\SSCC$ is
materialized with the help of the rewriting modulo
SMT~\cite{rocha-rewsmtjlamp-2017} approach, with constraints being
quantifier-free formulas over Boolean, integer, and real valued shared
variables, and information entailment queried as semantic inference
and automatically delivered by the SMT-based decision procedures. The
proposed automatic analysis of quantitative properties for $\SSCC$
relies on the $\PVESTA$ statistical model
checker~\cite{alturki-pvesta-2011}. Given a probabilistic rewrite
theory, such as the one presented in this paper, $\PVESTA$ is used to
simulate its execution while automatically evaluating expected values
of any numerical expression or path expression encoded in the
$\QUATEX$ language~\cite{agha-pmaude-2006}. This is achieved
automatically by performing enough Monte Carlo simulations to meet an
error threshold and still be meaningful for statistical inference. The
proposed approach is used to compute the expected execution time of a
reactive system. Of course, many other quantitative measures could be
computed by following an approach similar to the one presented in this
paper.

This work can be seen as yet another interesting use of rewriting
logic as a semantic framework. The support in rewriting logic for
real-time systems~\cite{olveczky-realtime-2002}, probabilistic
systems~\cite{agha-pmaude-2006}, and open
systems~\cite{rocha-rewsmtjlamp-2017}, make of the $\SSCC$ stochastic
and spatial rewriting logic semantics a symbolic and fully executable
specification in Maude~\cite{clavel-maudebook-2007} for constraint
systems exhibiting discrete and dense linear timing constraints. The
executable semantics can be used to specify and simulate the exchange
of information in a very general setting and perform automated
quantitative analysis for a broad class of reactive systems. All in
all, the $\SSCC$ semantics contributed by this work could become an
important test bed to formally model and assess quantitative
attributes of today's reactive systems.

Finally, this paper is a substantial extension of the conference
paper~\cite{ramirez-rtsccpwrla-2018} in the following ways:

\begin{itemize}
\item The $\SSCC$ executable rewriting logic semantics supports
  non-deterministic execution with exclusive (one process among many
  is chosen) and independent (some processes among many are chosen)
  choice, complementing the previous development in which only
  non-deterministic parallel execution was present.
\item The notion of processes with stochastic duration is also
  developed in this work; previously, processes could only be assigned a
  constant duration.
\item In terms of formal analysis, the semantics proposed in this work
  can be used for probabilistic simulation and quantitative analysis,
  extending the reachability and LTL model checking analysis
  previously offered in~\cite{ramirez-rtsccpwrla-2018}.
\item Extensive experimentation has been developed on a case study
  where a process randomly searches for specific information, encoded
  as a constraint, through a hierarchy of agents' spaces. Moreover,
  probabilistic simulation examples have been added.
\end{itemize}

\paragraph{Outline}
The rest of the paper is organized as
follows. Section~\ref{sec.prelim} collects preliminary notions on
concurrent constraint programming, rewriting logic, rewriting modulo
SMT, and probabilistic rewrite theories. Section~\ref{sec.rtsccs}
presents stochastic and spatial concurrent constraint programming and
Section~\ref{sec.rew} its rewriting logic semantics.
Section~\ref{sec.reach} presents some examples on probabilistic
simulation with the rewriting logic semantics and
Section~\ref{sec.case} presents a case study on quantitative
analysis. Section~\ref{sec.related} presents related work and
Section~\ref{sec.concl} concludes the paper. \ref{a.appendix} contains
the complete $\SSCC$ specification presented in Section~\ref{sec.rew}
and the code of the experiments presented in Section~\ref{sec.reach}.

\section{Preliminaries}
\label{sec.prelim}

The development of this paper relies on notions of process calculi,
concurrent constraint programming, and rewriting logic; all within the
general scope of concurrency theory. Concurrency theory is the field
of theoretical computer science concerned with the fundamental aspects
of systems consisting of multiple computing agents that interact among
each other. This covers a vast variety of systems including reactive
systems.

\subsection{Concurrent Constraint Programming and Constraint Systems}

Concurrent constraint programming
($\CCP$)~\cite{saraswat-ccp-1990,saraswat-ccpbook-1993,saraswat-ccpsem-1991}
(see a survey in \cite{olarte-emergmodels-2013}) is a model for
concurrency that combines the traditional operational view of process
calculi with a {\it declarative} view based on logic. This allows
$\CCP$ to benefit from the large set of reasoning techniques of both
process calculi and logic. Under this paradigm, the conception of
\emph{store as valuation} in the von Neumann model is replaced by the
notion of \emph{store as constraint} and processes are seen as
information transducers.

The $\CCP$ model of computation makes use of \emph{ask} and
\emph{tell} operations instead of the classical read and write. An ask
operation tests if a given piece of information (i.e., a constraint as
in $temperature > 23$) can be deduced from the store.  The tell
operations post constraints in the store, thus augmenting/refining the
information in it.  A fundamental issue in $\CCP$ is then the
specification of systems by means of constraints that represent
partial information about certain variables. The state of the system
is specified by the store (i.e., a constraint) that is monotonically
refined by processes adding new information.
  
The basic constructs (processes) in \emph{$\CCP$} are: (1) the ${\it
  tell}(c)$ agent, which posts the constraint $c$ to the store, making
it available to the other processes. Once a constraint is added, it
cannot be removed from the store (i.e., the store grows
monotonically). And (2), the ask process $c\rightarrow P$, which
queries if $c$ can be deduced from the information in the current
store; if so, the agent behaves like $P$, otherwise, it remains
blocked until more information is added to the store. In this way, ask
processes define a reactive synchronization mechanism based on
entailment of constraints. A basic $\CCP$ process language usually
adds \emph{parallel composition} ($P\parallel Q$) combining processes
concurrently, a \emph{hiding} operator for local variable definition,
and potential infinite computation by means of recursion or
replication. This notion of parallelism will be revisited in the light
of probabilistic choice.

The $\CCP$ model is parametric in a \emph{constraint system} ($\CS$)
specifying the structure and interdependencies of the partial
information that processes can query (\emph{ask}) and post
(\emph{tell}) in the \emph{shared store}. The notion of constraint
system can be given by using first-order logic. Given a signature
$\Sigma$ and a first-order theory $\Delta$ over $\Sigma$, constraints
can be thought of as first-order formulae over $\Sigma$. The (binary)
entailment relation $\vdash$ over constraints is defined for any pair
of constraints $c$ and $d$ by $c\entails d$ iff the implication $c
\Rightarrow d$ is valid in $\Delta$.

An algebra-based representation of $\CS$ is used in the present work.

\begin{definition}[Constraint Systems]
\label{def:cs}
	A constraint system $(\CS)$ $\mathbf{C}$ is a complete algebraic 
	lattice $(\Con, \sqsubseteq)$. The elements of $\Con$ are called 
	\emph{constraints}. The symbols $\sqcup$, $\true$ and $\false$ will 
	be used to denote the least upper bound $($lub$)$ operation, the
	bottom, and the top element of $\mathbf{C},$ respectively.
\end{definition}

\noindent In Definition~\ref{def:cs}, a $\CS$ is characterized as a
\emph{complete algebraic lattice}. The elements of the lattice, the
\emph{constraints}, represent (partial) information. A constraint $c$
can be viewed as an \emph{assertion} (or a \emph{proposition}). The
lattice order $\sqsubseteq$ is meant to capture entailment of
information: $\cleq{d}{c}$, alternatively written $\cgeq{c}{d}$, means
that the assertion $c$ represents as much information as $d$.  Thus
$\cleq{d}{c}$ may be interpreted as saying that $c\entails d$ or that
$d$ can be \emph{derived} from $c$. The \emph{least upper bound
$($lub$)$} operator $\sqcup$ represents join of information and thus
$\join{c}{d}$ is the least element in the underlying lattice above $c$
and $d$, asserting that both $c$ and $d$ hold. The top element
represents the lub of all, possibly inconsistent, information, hence
it is referred to as $\false$. The bottom element $\true$ represents
the empty information.

\subsection{Order-sorted Rewriting Logic in a Nutshell}

Rewriting logic~\cite{meseguer-rltcs-1992} is a general semantic
framework that unifies a wide range of models of concurrency. Language
specifications can be executed in Maude~\cite{clavel-maudebook-2007},
a high-performance rewriting logic implementation and benefit from a
wide set of formal analysis tools available to it, such as an LTL
model checker and an inductive theorem prover. The reader is referred
to~\cite{meseguer-rltcs-1992,clavel-maudebook-2007,meseguer-rlsproject-2013,rusu-rewt-2016}
for an in-depth treatment of the topics discussed next.

\subsubsection{Rewriting Logic}
A {\em rewriting logic specification} or {\em rewrite theory} is a tuple
$\rcal = (\Sigma, E \uplus B, R)$ where:
\begin{itemize}
\item $(\Sigma, E\uplus B)$ is an order-sorted equational theory with
  $\Sigma = (S,\leq,F)$ a signature with finite poset of sorts
  $(S,\leq)$ and a set of function symbols $F$ typed with sorts in
  $S$; $E$ is a set of $\Sigma$-equations, which are universally
  quantified Horn clauses with atoms that are $\Sigma$-equations $t=u$
  with $t,u$ terms of the same sort; $B$ is a set of structural axioms
  --- disjoint from the set of equations $E$ ---
  (e.g., associativity, commutativity, identity) such that there
  exists a matching algorithm modulo $B$ producing a finite number of
  $B$-matching substitutions or failing otherwise; and
\item $R$ a set of universally quantified conditional rewrite rules
  of the form 
  \[\crl{t}{u}{\bigwedge_i \phi_i}\]
  where $t,u$ are $\Sigma$-terms of the same sort and each $\phi_i$ is
  a $\Sigma$-equality.
\end{itemize}
Given $X = \{X_s\}_{s\in S}$, an $S$-indexed family of disjoint
variable sets with each $X_s$ countably infinite, the {\em set of
  terms of sort $s$} and the {\em set of ground terms of sort $s$} are
denoted, respectively, by $T_\Sigma(X)_s$ and $T_{\Sigma,s}$;
similarly, $T_\Sigma(X)$ and $T_\Sigma$ denote, respectively, the set
of terms and the set of ground terms. The expressions
$\tcal_\Sigma(X)$ and $\tcal_\Sigma$ denote the corresponding
order-sorted $\Sigma$-term algebras.  All order-sorted signatures are
assumed {\em preregular}~\cite{goguen-ordersorted-1992}, i.e., each
$\Sigma$-term $t$ has a unique {\em least sort} $\ls{t} \in S$ s.t. $t
\in T_\Sigma(X)_{\ls{t}}$. It is also assumed that $\Sigma$ has {\em
  nonempty sorts}, i.e., $T_{\Sigma,s}\neq \emptyset$ for each $s\in
S$. Many-sorted equational logic is the special case of order-sorted
equational logic when the subsort relation $\leq$ is restricted to be
the identity relation over the sorts.

An equational theory $\ecal = (\Sigma,E \uplus B)$ induces the
congruence relation $=_\ecal$ on $T_\Sigma(X)$ (or simply $=_{E \uplus
  B}$) defined for $t,u \in T_\Sigma(X)$ by $t =_\ecal u$ if and only
if $\ecal \ded \eq{t}{u}$, where $\ecal \ded \eq{t}{u}$ denotes
$\ecal$-provability by the deduction rules for order-sorted equational
logic in~\cite{meseguer-membership-1998}. For the purpose of this
paper, such inference rules, which are analogous to those of
many-sorted equational logic, are even simpler thanks to the
assumption that $\Sigma$ has nonempty sorts, which makes unnecessary
the explicit treatment of universal quantifiers. The expressions
$\tcal_{\ecal}(X)$ and $\tcal_\ecal$ (also written
$\tcal_{\Sigma/E\uplus B}(X)$ and $\tcal_{\Sigma/E \uplus B}$) denote
the quotient algebras induced by $=_\ecal$ on the term algebras
$\tcal_\Sigma(X)$ and $\tcal_\Sigma$, respectively; $\tcal_{\Sigma/E
  \uplus B}$ is called the {\em initial algebra} of $(\Sigma,E \uplus
B)$.


A rewrite theory $\rcal = (\Sigma, E \uplus B, R)$ induces a rewrite
relation $\rews_{\rcal}$ on $T_{\Sigma}(X)$ (sometimes denoted also as
$\rews_{R/E \uplus B}$) defined for every $t,u \in T_\Sigma(X)$ by $t
\rews_\rcal u$ if and only if there is a rule $(\ccrl{l}{r}{\phi}) \in
R$ and a substitution $\func{\theta}{X}{T_\Sigma(X)}$ satisfying $t
=_{E\uplus B} l\theta$, $u =_{E \uplus B} r\theta$, and $E \vdash
\phi\theta$. The tuple $\tcal_\rcal = (\tcal_{\Sigma/{E \uplus
    B}},\rews_\rcal^*)$ is called the {\em initial reachability model
  of $\rcal$}~\cite{bruni-semantics-2006}.

\subsubsection{Admissible Rewrite Theories}
Appropriate requirements are needed to make an equational theory
$\ecal$ {\em admissible}, i.e., {\em executable} in rewriting
languages such as Maude~\cite{clavel-maudebook-2007}.  In this paper, it is
assumed that the equations $E$ can be oriented into a set of (possibly
conditional) {sort-decreasing}, {operationally terminating}, and
{confluent} rewrite rules $\overrightarrow{E}$ modulo $B$ 
(denoted by $\rews_{E/B}$ and equivalent to $=_{B} \rews_E
=_{B}$). The rewrite system $\overrightarrow{E}$ is
{\em sort decreasing} modulo $B$ if and only if for each
 $(\crl{t}{u}{\cond}) \in \overrightarrow{E}$ and
substitution $\theta$, $\ls{t\theta} \geq \ls{u\theta}$ if
$(\Sigma,B,\overrightarrow{E}) \ded \cond\theta$.  The system
$\overrightarrow{E}$ is {\em operationally terminating} modulo
$B$~\cite{duran-operterm-2008} if and only if there is no infinite
well-formed proof tree in $(\Sigma,B,\overrightarrow{E})$
(see~\cite{lucas-ordersorted-2009} for terminology and details).  
Furthermore, $\overrightarrow{E}$ is {\em confluent} modulo $B$ if 
and only if for all $t,t_1,t_2 \in T_\Sigma(X)$, if $t \rews^*_{E/B} t_1$ 
and $t \rews^*_{E/B} t_2$, then there is $u \in T_\Sigma(X)$ such that $t_1
\rews^*_{E/B} u$ and $t_2 \rews^*_{E/B} u$.  The term $\can{t}{E/B}
\in T_\Sigma(X)$ denotes the {\em $E$-canonical form} of $t$ modulo
$B$ so that $t \rews_{E/B}^* \can{t}{E/B}$ and $\can{t}{E / B}$ cannot
be further reduced by $\rews_{E/B}$. Under sort-decreasingness,
operational termination, and confluence, the term $\can{t}{E/B}$ is
unique up to $B$-equality.

For a rewrite theory $\rcal$, the rewrite relation $\rews_\rcal$ is
undecidable in general, even if its underlying equational theory is
admissible, unless conditions such as
coherence~\cite{viry-coherence-2002} are given (i.e, whenever
rewriting with $\rews_{R/E \cup B}$ can be decomposed into rewriting
with $\rews_{E/B}$ and $\rews_{R/B}$). A key goal
of~\cite{rocha-rewsmtjlamp-2017} was to make the relation
$\rews_\rcal$ both decidable and symbolically executable when $E$
decomposes as $E_\lang \uplus B_\nlang$, representing a built-in
theory $E_\lang$ for which formula satisfiability is decidable and
$B_\nlang$ has a matching algorithm.

\subsubsection{Rewriting Logic Semantics}
The rewriting logic semantics of a language $\lcal$ is a rewrite
theory $\rcal_\lcal = (\Sigma_\lcal,E_\lcal \uplus B_\lcal, R_\lcal)$
where $\rews_{\rcal_\lcal}$ provides a step-by-step formal description
of $\lcal$'s {\em observable} run-to-completion mechanisms. The
conceptual distinction between equations and rules in $\rcal_\lcal$
has important consequences that are captured by rewriting logic's {\em
  abstraction dial}~\cite{meseguer-rlsproject-2013}. Setting the level of
abstraction in which all the interleaving behavior of evaluations in
$\lcal$ is observable, corresponds to the special case in which the
dial is turned down to its minimum position by having $E_\lcal \uplus
B_\lcal = \emptyset$. The abstraction dial can also be turned up to
its maximal position as the special case in which $R_\lcal =
\emptyset$, thus obtaining an equational semantics of $\lcal$ without
observable transitions. The rewriting logic semantics presented in
this paper is {\em faithful} in the sense that such an abstraction
dial is set at a position that exactly captures the interleaving
behavior of the concurrency model.

\subsubsection{Probabilistic Rewrite Theories}


In a \emph{probabilistic rewrite theory}~\cite{agha-pmaude-2006},
rewrite rules can have the more general form
\[\pcrl{l(\tb{x})}{r(\tb{x},\tb{y})}{\phi(\tb{x})}{\tb{y} \;{:=}\; \pi(\tb{x})}\]
Because the pattern $r(\tb{x},\tb{y})$ on the right-hand side
may have new variables $\tb{y}$, the next state specified by such a
rule is not uniquely determined: it depends on the choice of an
additional substitution $\rho$ for the variables $\tb{y}$. In this
case, the choice of $\rho$ is made according to the family of
probability functions $\pi_\theta$: one for each matching substitution
$\theta$ of the variables $\tb{x}$. Therefore, a probabilistic
rewrite theory can express both non-deterministic and probabilistic
behavior of a concurrent system. At any given point of execution of a
probabilistic rewrite theory many different rules can be enabled. Once
a matching substitution $\theta$ has been chosen for one of these
rules, the choice of the substitution $\rho$ is made probabilistically
according to the probability distribution function $\pi_\theta$.

\subsubsection{Real time}

Time sampling strategies offer alternatives to assign the time that a
rewrite step needs to be applied. For instance, the maximal time
sampling strategy advances time by the maximum possible time elapse
and tries to advance time by a user-given time value in tick rules
having other forms.  There are two different kinds of tick rule
applications that the maximal strategy can treat: (i) ticks from
states from which time can only advance up to a certain maximal time,
and (ii) ticks from states from which time can advance by any
amount. Here, the tick rule is the second one and the maximal time
sampling strategy handles it by advancing time by a user-given time
value.

A time-robust system is one where from any given state time can
advance either by: (i) any amount, (ii) any amount up to (and including)
a specific instant in time, or (iii) not at all. Advancing time is not
affected unless in a specific state time is advanced all the way to the 
specific bound in time given in (ii). An instantaneous rewrite rule can
only be applied at specific times, namely, when the system has advanced
time by the maximal possible amount.

A time-robust system may have \emph{Zeno paths}, those are paths where
the sum of the durations of an infinite number of tick steps is
bounded.  It is necesary to differentiate between Zeno paths forced by
the specification and Zeno paths that are due to bad choices in the
tick increments.  The intuition in the second type of Zeno behavior
does not reflect realistic behaviors in the system and therefore is
not simulated by the maximal time sampling strategy.  The paths of the
system that do not exhibit this unrealistic kind of Zeno behavior are
called \emph{timed fair paths}.

For systems satisfying time-robustness and tick-stabilizing properties
unbounded and time-bounded LTL (excluding the \emph{next} operator) model 
checking using the maximum time elapsed strategy is
complete~\cite{olveczky-acrealtime-2007}.

\subsubsection{Maude, PMaude, and $\PVESTA$}
\label{sec.prelim.pvesta}

Maude~\cite{clavel-maudebook-2007} is a language and system based on
rewriting logic.  It supports order-sorted equational and rewrite
theory specifications in \emph{functional} and \emph{system} modules,
respectively. Admissibility of functional and system modules can be
checked with the help of the \emph{Maude Formal Environment}
(MFE)~\cite{duran-mfetalcott-2011,duran-mfecalco-2011}, an executable
formal specification in Maude with tools to mechanically verify such
properties.  The MFE includes the Maude Termination Tool, the Maude
Sufficient Completeness Checker, the Church-Rosser Checker, the
Coherence Checker, and the Maude Inductive Theorem Prover.
All this tools are available at \url{http://maude.lcc.uma.es/MFE}.

PMaude~\cite{agha-pmaude-2006} is both a language for specifying
probabilistic rewrite theories and an extension of Maude supporting
the execution of such theories by discrete-event simulation. PMaude
can capture the dynamics of various elements of a system by stochastic
real-time: computation and message-passing between entities of a
system may take some positive real-valued time that can be distributed
according to some continuous probability distribution function. Time
associated to computation and message passing can also be zero,
indicating instant transitions and synchronous communication. In
general, PMaude supports discrete probabilistic choice as found in
discrete-time Markov chains and stochastic continuous-time as found in
continuous-time Markov chains.

A specification in PMaude without \emph{unquantified non-determinism}
is a key requirement for the statistical model checking analysis.
Intuitively, non-existence of unquantified non-determinism means that
non-de\-ter\-mi\-nis\-tic choice during the simulation of a
probabilistic rewrite theory $\rcal$ is exclusively due to
probabilistic choice and not to concurrent transitions firing
simultaneously at different parts of a system state. Under this
assumption (and the admissibility assumptions on $\rcal$), a one-step
computation with $\to_\rcal$ represents a single step in a
discrete-event simulation of a specification written in PMaude. For
details about one-step computation and sufficient conditions for
absence of unquantified non-determinism in PMaude, we refer the
interested reader to~\cite{agha-pmaude-2006}.  The rewriting logic
semantics $\textsf{sscc}$ in Section~\ref{sec.rew} is assumed to be
admissible. Freeness of unquantified non-determinism is achieved by
using a scheduler that is deterministic (in the sense that its
behavior is deterministic relative to a given seed for random number
generation).

Once a probabilistic system has been modeled in PMaude, various
quantitative properties of the system can be specified by using the
\emph{Quantitative Temporal Expressions} language
($\QUATEX$)~\cite{agha-pmaude-2006} and queried with the help of the
$\PVESTA$ statistical model checker~\cite{alturki-pvesta-2011}.  The
reader is referred to~\cite{agha-pmaude-2006} for additional details
about $\QUATEX$ syntax and semantics, how other logics such as the
\emph{Probabilistic Computation Tree Logic}
(PCTL)~\cite{hansson-pctl-1994} can be encoded in it, and the
mechanisms used by $\PVESTA$ for statistical evaluation of $\QUATEX$
expressions. Formally, given a probabilistic model $\mathcal{M}$, an
expectation QuaTEx formula of the form $\mathbf{E}[Exp]$ ---with $Exp$
a QuaTEx expression---, and bounds $\alpha$ and $\delta$, $\PVESTA$
approximates the value of $\mathbf{E}[Exp]$ within a $(1-\alpha)100\%$
confidence interval and with size at most $\delta$. This is done by
generating a large enough number $n$ of random sample values $x_1,
x_2, \dots, x_n$ of $Exp$, computed from $n$ independent Monte Carlo
simulations of $\mathcal{M}$~\cite{alturki-pvesta-2011}.

$\PVESTA$ is implemented in Java 1.6 and it is available
at~\url{http://maude.cs.uiuc.edu/tools/pvesta/}.

\subsection{SMT Solving}

Satisfiability Modulo Theories (SMT) studies methods for checking
satisfiability of first-order formulas in specific models. The SMT
problem is a decision problem for logical formulas with respect to
combinations of background theories expressed in classical first-order
logic with equality. An SMT instance is a formula $\phi$ (typically
quantifier free, but not necessarily) in first-order logic and a model
$\tcal$, with the goal of determining if $\phi$ is satisfiable in
$\tcal$.

In this work, the representation of the constraint system is based on
SMT solving technology. Given a many-sorted equational theory
$\ecal_\lang = (\Sigma_\lang,E_\lang)$ and a set of variables $X_\lang
\subseteq X$ over the sorts in $\Sigma_\lang$, the formulas under
consideration are in the set $\oqff{\Sigma_\lang}{X_\lang}$ of
quantifier-free $\Sigma_\lang$-formulas: each formula being a Boolean
combination of $\Sigma_\lang$-equation with variables in $X_\lang$
(i.e., atoms). The terms in $T_{\ecal_\lang}$ are called
\textit{built-ins} and represent the portion of the specification that
will be handled by the SMT solver (i.e., semantic data types). In this
setting, an SMT instance is a formula $\phi \in
\oqff{\Sigma_\lang}{X_\lang}$ and the initial algebra
$\tcal_{\ecal_\lang^{+}}$, where $\ecal_\lang^{+}$ is a
\textit{decidable extension} of $\ecal_\lang$ such that
\begin{align*}
  \phi
  \textnormal{ is satisfiable in $\tcal_{\ecal_\lang^+}$} \; \iff \; (\exists
  \func{\sigma}{X_\lang}{T_{\Sigma_\lang}})\; \tcal_{\ecal_\lang}\models\phi\sigma.
\end{align*}
Many decidable theories $\ecal_{\lang}^+$ of interest are supported by
SMT solvers satisfying this requirement
(see~\cite{rocha-rewsmtjlamp-2017} for details).  In this work, the
Maude alpha 118 release, which integrates
Yices2~\cite{dutertre-yices2-2014} and CVC4~\cite{barrett-cvc4-2011},
is used for reachability analysis with SMT constraints.

\section{Stochastic and Spatial Concurrent Constraint Systems}
\label{sec.rtsccs}

In this section we present the stochastic and spatial concurrent constraint
($\SSCC$) calculus and illustrate the main features of the language.

\subsection{Spatial Constraints}
\label{sec.sccp}

The authors of \cite{knight-sccp-2012} extended the notion of $\CS$ to
account for distributed and multi-agent scenarios where agents have
their own space for local information and computation. In  \cite{guzman:hal-01257113,guzman:hal-01328188,guzman:hal-01675010,haar:hal-01256984,guzman:hal-02172415}
$\CS$ are further extended to model mobile behaviour and reason about beliefs, lies, and group epistemic behaviour inspired by social networks. 

\paragraph{Locality and Nested Spaces}
Each agent $i$ has a \emph{space} function $\sfunc{\cdot}_i$ from
constraints to constraints (recall that constraints can be viewed as
assertions). Applying the space function  $\sfunc{\cdot}_i$  to a constraint $c$ gives us a constraint $\sfunc{c}_i$ that can be interpreted as an
assertion stating that $c$ is a piece of information that resides
\emph{within a space attributed to agent} $i$. An alternative
\emph{epistemic interpretation} of $\sfunc{c}_i$ is an assertion
stating that agent $i$ \emph{believes} $c$ or that $c$ holds within
the space of agent $i$ (but it may or may not hold elsewhere). Both
interpretations convey the idea that $c$ is local to agent
$i$. Following this intuition, the assertion $\sfunc{\sfunc{c}_j}_i$
is a hierarchical spatial specification stating that $c$ holds within
the local space the agent $i$ attributes to agent $j$. Nesting of
spaces such as in $\sfunc{\sfunc{\cdots\sfunc{c}_{i_m}
    \cdots}_{i_2}}_{i_1}$ can be of any depth.

\paragraph{Parallel Spaces}
A constraint of the form $\join{\sfunc{c}_i}{\sfunc{d}_j}$ can be seen
as an assertion specifying that $c$ and $d$ hold within two
\emph{parallel/neighboring} spaces that belong to agents $i$ and
$j$. From a computational/concurrency point of view, it is possible to
think of $\sqcup$ as parallel composition; from a logic point of view,
$\sqcup$ corresponds to conjunction.

The notion of an $n$-agent spatial constraint system is formalized in
Definition~\ref{def.sccp.scs}.
\begin{definition}[Spatial Constraint system \cite{knight-sccp-2012}]
  \label{def.sccp.scs}
  An $n$-agent \emph{spatial constraint system ($n$-$\SCS$)}
  ${\bf C}$ is a $\CS$ $(\Con, \sqsubseteq)$ equipped with $n$ self-maps
  $\sfuncs$ over its set of constraints ${\Con}$ satisfying for each
  function $\sfunc{\cdot}_i:\Con \rightarrow \Con$:
  \begin{description}
    \item[S.1] $\sfunc{\true}_i = \true$, \text{and}
    \item[S.2] $\sfunc{\join{c}{d}}_i = \join{\sfunc{c}_i}
    {\sfunc{d}_i} \ \ \text{ for each } c,d \in \Con.$
  \end{description}
\end{definition}
Property S.1 in Definition \ref{def.sccp.scs} requires space functions
to be strict maps (i.e., bottom preserving) where an empty local space
amounts to having no knowledge. Property S.2 states that space
functions preserve (finite) lubs, and also allows to join and
distribute the local information of any agent $i.$ Henceforth, given
an $n$-$\SCS$ ${\bf C}$, each $\sfunc{\cdot}_i$ is thought as the
\emph{space} (or space function) of the agent $i$ in ${\bf C}$. The
tuple $({\Con},\sqsubseteq,\sfuncs)$ denotes the corresponding
$n$-$\SCS$ with space functions $\sfuncs.$ Components of an $n$-$\SCS$
tuple shall be omitted when they are unnecessary or clear from the
context. When $n$ is unimportant, $n$-$\SCS$ is simply written as
$\SCS$.

Mobility plays a key role in distributed systems.  Following the
algebraic approach, under certain conditions it is possible to provide each agent $i$ with an
\emph{extrusion} function $\uparrow_i : \Con \rightarrow \Con$
~\cite{haar:hal-01256984,guzman:hal-01257113}. The expression $\uparrow_i c$ within a space
context $\sfunc{\cdot}_i$ means that $c$ must be posted
outside of agent's $i$ space. 

More precisely, given a space function $\sfunc{\cdot}_i$, the extrusion function $\uparrow_i$ of agent $i$ is  
the  \emph{right inverse}  of  $\sfunc{\cdot}_i$. Such function exists if and only if  $\sfunc{\cdot}_i$ is surjective \cite{haar:hal-01256984}. By right inverse of  $\sfunc{\cdot}_i$ we mean a function $\efunc{_i}:\Con \rightarrow \Con $ such that  $\sfunc{\ \uparrow_i c \ }_i = c$. The computational interpretation of $\uparrow_i c$ is that of a process being able to extrude any $c$ from the space  $\sfunc{\cdot}_i.$ The extruded information $c$ may not necessarily be part of the information residing in the space of agent $i$. For example, using properties of space and extrusion
functions we shall see that $\sfunc{\ d \ \sqcup \uparrow_i c }_i = \sfunc{ d  }_i  \sqcup c$ specifying that $c$ is extruded (while $d$ is still in the space of $i$).  The extruded $c$  could be inconsistent with $d$ (i.e., $c \sqcup d =\false$), it could be related to $d$ (e.g., $c  \sqsubseteq d$), or simply unrelated to $d$. From an epistemic perspective, we can use $\uparrow_i$  to express  \emph{utterances} by agent $i$ and such utterances could be intentional lies (i.e., inconsistent with their beliefs), informed opinions (i.e., derived from the beliefs), or simply arbitrary statements (i.e., unrelated to their beliefs). One can then think of extrusion/utterance  as the \emph{right inverse} of space/belief.  

We can now recall the notion of spatial constraint system with extrusion.

\begin{definition}[Spatial Constraint System with Extrusion~\cite{haar:hal-01256984,guzman:hal-01257113}]
  \label{def.sccp.scse}
  An \emph{$n$-agent spatial constraint system with extrusion
    ($n$-$\SCSE$) is an $n$-$\SCS$ ${\bf C}$ equipped with $n$
    self-maps $\uparrow_1, \dots , \uparrow_n$ over $\Con$, written
    $({\bf C}, \uparrow_1, \dots , \uparrow_n)$}, such that each
  $\uparrow_i$ is the right inverse of $\sfunc{\cdot}_i$.
\end{definition}

\paragraph{Agent Views} Let us recall the notion of agent view. 

\begin{definition}[Agent View \cite{knight-sccp-2012}]\label{view} The agent $i$'s \emph{view} of $c$, $\prj ci$, is given by $\prj ci=\bigsqcup\{d\;|\;\sfunc{d}_i \sqsubseteq c\}$.
\end{definition}
Intuitively, $\prj ci$ represents all the information the agent $i$ may see or have in $c$.  For example if $c=\sfunc{d}_i  \sqcup \sfunc{e}_j $
then agent $i$ sees $d$, so $d \sqsubseteq \prj ci$. 

\subsection{Spatial Concurrent Constraint Programming with Probabilistic Choice}

This section presents the syntax of $\SSCC$ and main intuition behind
its constructs. The operational semantics of $\SSCC$ will be given in Section \ref{sec.rew}.

\begin{definition}[$\PTSCCP$\ Processes]\label{def.sccp.sccpe.syntax}
  Let ${\bf C}=({\Con},\sqsubseteq)$ be a constraint system, $A$ a
  set of $n$-agents, and $V$ an infinite countable set of
  variables. Let $(\textbf{C}, \sfuncs, \efuncs)$ be an $n$-$\SCSE$
  and consider the following EBNF-like syntax:
  \vspace{-0.15cm}
  \begin{align*} P & \; ::= \;
  \Stop\;\Big|\; 
  \tell(c) \;\Big|\; 
  \ask(c)\rightarrow P \;\Big|\; 
  P \parallel P \;\Big|\; 
  \K i P \;\Big|\; 
  \extr{i}{P} \;\Big|\;
  x \;\Big|\; 
  \mu x.P \;\Big|\; 
  \bigoplus_j (P,q)_j \;\Big|\; 
  \bigodot_j (P,q)_j 
  \end{align*}
  \noindent where $c \in \Con$, $i \in A$, $x \in V$, $j$ belongs
  to a finite set of indexes $J$, and $q \in [0, 1]$.
  An expression $P$ in the above syntax is a \emph{process}
  if and only if every variable $x$ in $P$ occurs in the scope
  of an expression of the form $\mu x.P$. The set of processes
  of $\PTSCCP$ is denoted by $\textit{Proc}$.
\end{definition}

\noindent The $\PTSCCP$ calculus can be thought of as a
\emph{shared-spaces} model of computation. Each agent $i \in A$ has a
computational space of the form ${\K i \cdot}$ possibly containing
processes and other agents' spaces.  The basic constructs of $\PTSCCP$
are {\em tell}, {\em ask}, and {\em parallel} composition, and they
are defined as in standard $\CCP$~\cite{saraswat-ccpsem-1991}.  A
process $\tell(c)$ running in an agent $i \in A$ adds $c$ to its local
store $s_i$, making it available to other processes in the same
space. This addition, represented as $s_i \sqcup c$, is performed even
if the resulting constraint is inconsistent.  The process
$\ask(c)\rightarrow P$ running in space $i$ may execute $P$ if $c$ is
entailed by $s_i$, i.e., $c \sqsubseteq s_i$. The process $P \parallel
Q$ specifies the {\it parallel execution} of processes $P$ and $Q$;
given $I=\{ i_1,\ldots,i_m \}$, the expression $\prod_{i
  \in I} P_i$ is used as a shorthand for $P_{i_1} \parallel \ldots
\parallel P_{i_m}$.  A construction of the form ${\K iP}$ denotes a
process $P$ running within the agent $i$'s space. Any information that
$P$ produces is available to processes that lie within the same
space. The process $\extr{i}{P}$ denotes that process $P$ runs outside
the space of agent $i$ and the information posted by $P$ resides in
the store of the parent of agent $i$. Unbounded behaviour is specified using recursive definitions of the form $\mu x.P$ whose behaviour is that of $P[\mu x.P/x]$, i.e., $P$ with every free occurrence of $x$ replaced with $\mu x.P.$ We assume that 
recursion is \emph{ask guarded}: i.e., for every $\mu x. P$, each occurrence of  $x$ in $P$ occurs under the scope of an ask process. For simplicity we assume an implicit  ``$\ask(\true)\rightarrow"$ in unguarded occurrences of $X$. 


The last two processes represent exclusive and independent
probabilistic choice, which are part of the main contribution of this
work.  The pair $(P, q)_j$ is a shorthand for $(P_j, q_j)$ and
represents a process $P_j$ with probability $q_j$ to be scheduled for
execution. The probability that $P_j$ is not scheduled is given by $1
- q_j$. The $\bigoplus_j\ (P,q)_j$ process represents the exclusive
choice of some process $P_k$ with probability $q_k$ where $k \in J$
and $\sum\limits_{j \in J} q_j = 1$.  The $\bigodot_j\ (P,q)_j$
process represents the independent choice (possibly none) of some
processes $P_{j_1}, P_{j_2}, \dots, P_{j_k}$ each one with probability
$q_{j_1}, q_{j_2}, \dots, q_{j_k}$, respectively, where $j_h \in K$
(with $1 \leq h \leq k$), for some (possibly empty) $K \subseteq
J$. Once a choice is made, $\bigodot_j\ (P,q)_j$  evolves to the parallel composition of the chosen processes $\prod_{k \in K} P_k$
and the remaining processes (i.e., those indexed by $J \setminus K$)
are precluded.
 
\begin{example}\label{exa.sccp.spatial}
  Consider the processes $P = \tell(c)$ and $Q=\ask(c)\rightarrow
  \tell(d)$.
  \begin{itemize}
  \item The process ${\K i P } \parallel {\K i {Q} }$, by the above
    intuitions, have the effect that the constraints $c$ and $d$ are
    added to store of agent $i$.
  \item A similar behavior is achieved by the process ${\K i {P
      \parallel Q}}$, which also produces $c \sqcup d$ in the store of
    agent $i$ (note that $\sfunc{c \sqcup d}_i$ is equivalent to
    $\sfunc{c}_i\sqcup \sfunc{d}_i$ by Property S.2 in Definition
    \ref{def.sccp.scs}).
  \item In contrast, the process ${\K j {P} } \parallel {\K i {Q}}$,
    with $i\neq j$, does not necessarily add $d$ to the space of agent
    $i$ because $c$ is not made available for agent $i$; likewise in
    $P \parallel {\K i {Q}}$, $d$ is not added to the space of agent
    $i$.

  \item Consider ${\K i {P \parallel {\K j {\extr{j}{Q}}}}}$. In this
    case, because of extrusion, both $c$ and $d$ will be added to
    store of agent $i$. However, ${\K j {\K i {P \parallel
          {\extr{i}{Q}}}}}$ with $i\neq j$, adds $c$ to  the space of
    agent $i$ within the space of agent $j$, but $c$ is not made available for agent $j$, therefore $d$ could not be added to its space. Note that in ${\K i P} \parallel {\K j {\extr{i}{Q}}}$, the
    constraint $c$ is added to the space of agent $i$, but since $Q$
    cannot be extruded in ${\K j {\extr{i}{Q}}}$, $d$ is not added
    neither to the space of $i$ nor $j$.

    \item Consider the process $(\K i P, q_1) \odot (\K i Q, q_2)$ and
      a random sampling $p = 0.5$. If $q_1 = q_2 =0.7$, the store of
      agent $i$ is modified with $c$ and $d$ because $q_1 \geq p$ and
      $q_2 \geq p$. If $q_1 = 0.7$ and $q_2 = 0.4$ however, $c$ is
      added to the store but $Q$ cannot be executed within the space
      of agent $i$ because $q_1 \geq p$ but $q_2 \not\geq p$.

    \item For exclusive choice, consider $(\K i P, q_1) \oplus(\K j P,
      q_2)$ and a random sampling $p = 0.5$. In this case, the
      exclusive choice in $(\K i P, q_1) \oplus(\K j P, q_2)$
      determines which agent, either $i$ or $j$, is going to add to
      its store, but only one of them is able to add such a
      constraint. If $q_1 = 0.7$ and $q_2 = 0.3$, $c$ is added to the
      store of agent $i$.
  \end{itemize}
\end{example}

\subsection{Configurations}
As usual configuration are used to represent the state of the system.
 A configuration is a pair of the form $\pairccp{P}{c} \in
Proc \times \Con$, where $P$ is a process and $c$ is the spatial
distribution of information available to it.

\begin{example}
  Consider the constraint $d$ below and its tree-like structure depicted
  in Figure~\ref{fig.sccp.tree}.

  \[
    d \defsymbol
    (y=1) \sqcup
    \sfunc{x=3 \sqcup\sfunc{y=3}_j}_i \sqcup
    \sfunc{x>0 \sqcup\sfunc{x=42}_k \sqcup \sfunc{x<42}_j}_j \sqcup
    \sfunc{y>0}_k.
  \]

  Each node in such a tree corresponds to the information (constraint)
  contained in an agent's space. Edges define the spatial hierarchy
  of agents. For example, the configuration
  $\Conf{\sfunc{\askp{x=42}{P}}_j}{d}$ is a deadlock, while
  $\Conf{\sfunc{\sfunc{\askp{x=42}{P}}_k}_j}{d}$  can evolve to 
  $\Conf{\sfunc{\sfunc{P}_k}_j}{d}$ since
  $\cleq{x=42}{\prj{{\prj dj}}k}$ (see Def.\ref{view}).
\end{example}
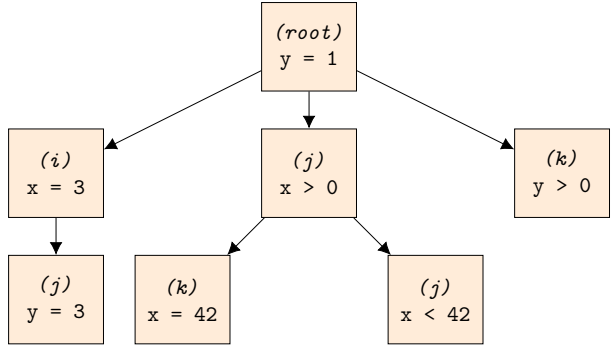
\begin{figure}[thbp] 
  \centering
  \resizebox{8cm}{!}{%
    \begin{tikzpicture}[node distance=2cm,
    every node/.style={fill=white, font=\sffamily}, align=center]
    \node (root)  [store]                
    {\textit{(root)}\\y = 1};
    \node (i)   [store, below of=root, left of=root, xshift=-2cm]
    {\textit{(i)}\\x = 3};
    \node (j)   [store, below of=root]  
    {\textit{(j)}\\x > 0};
    \node (k)   [store, below of=root, right of=root, xshift=2cm]  
    {\textit{(k)}\\y > 0};           
    \node (ji)    [store, below of=i] 
    {\textit{(j)}\\y = 3};
    \node (kj)    [store, below of=j, left of=j] 
    {\textit{(k)}\\x = 42};
    \node (jj)    [store, below of=j, right of=j] 
    {\textit{(j)}\\x < 42};         
    \draw[->]   (root) -- (i);
    \draw[->]   (root) -- (j);
    \draw[->]   (root) -- (k);     
    \draw[->]      (i) -- (ji);
    \draw[->]      (j) -- (kj);
    \draw[->]      (j) -- (jj);      
    \end{tikzpicture}}
  \caption{A spatial hierarchy of processes.}
  \label{fig.sccp.tree}
\end{figure}

\subsection{Timed Processes in \PTSCCP}
\label{tpro}

In these systems time is conceptually divided into discrete or continuous
intervals. In a time interval, the processes may receive a stimulus from
the environment and react computing and responding to it.
Time can be thought of as a sequence of time slots. The processes make their
internal transitions in a given time unit. When the current unit ends the
processes that are pending to finish their transitions are passed to the
next time unit. In this system there is no time limit, i.e., the processes
make their internal transitions until no further transition can be done.
Note that the processes remained at the end of the execution are only $ask$
processes.
They are the processes for which the store may not have enough information to
derive their condition.

Modeling real-time involves a scheduler to manage the processes preemption
based on the time that they require to complete their transitions. Thereby,
a process makes its transitions when no other process has a lower time out.

This approach uses a time-unit $T$ in which transitions can be done.
If a process cannot be reduced because there is not enough information, it
waits to complete its reduction, if it is possible, in the next time-unit.
Accessing the store has a time penalty, namely, \emph{tell} processes have
a cost for modifying the store and \emph{ask} processes have a for querying
the store. Mobility implies a cost for changing spaces, for $\K i \cdot$
this cost represents the time of getting in the space of agent $i$, and
for $\extr{i}{\cdot}$ is the time of leaving the space of agent $i$.
This time is given by probability distribution functions ($\alpha, \mu, \phi$ and $\rho$, respectively) in each space and
for each one of the processes, therefore each process could have different
times within each space.


%
\begin{example}\label{exa.rtsccp.time}
	As an example, consider the tree-like structures depicted in
	Figure~\ref{fig.rtsccs.rtsc}. They correspond to hierarchical
	computational spaces of, e.g., virtual containerization (i.e., virtual
	machines inside other virtual machines). Each one of these spaces is
	endowed with an agent identifier (either \cde{root} or a natural
	number) and a local store (i.e., a constraint), and the processes can
	be executed and spawned concurrently inside any space, with the
	potential to traverse the structure, querying and posting information
	locally, and even creating new spaces. The $\SCCP$ calculus enables
	the formal modeling of such scenarios and of transitions that can lead
	from an initial system state (e.g.,
	Figure~\ref{fig.rtsccs.rtsc-initial}) to a final state (e.g.,
	Figure~\ref{fig.rtsccs.rtsc-final}) by means of an operational
	semantics~\cite{knight-sccp-2012}.
	
	The initial state is represented by the constraint $d$ specifying the
  store of each space. The final state is reached after execution of the configuration $C$: 
	\[ d \defsymbol (W=9) \sqcup \sfunc{X \geq11}_0 \sqcup \sfunc{true
		\sqcup\sfunc{Y>5}_0}_1 \sqcup\sfunc{true}_2, \]
	\[ C\defsymbol \conf{{\K 0 {\askp{X>2}{\extr{0}{\K 1 {\K 0 
		{\tellp{Y<10}}}}}}} \parallel {\K 2 {\tellp{Z \neq 10}}} }
     {root}{d}{0}. \] The configuration $C$ reduces the container
     system $d$ to the state in Figure~\ref{fig.rtsccs.rtsc-final}
     using the time functions:
	\[\alpha \defsymbol \{(root,0.1),(0.root,0.15),(1.root,0.15),(2.root,0.15),
		(0.1.root,0.2)\},\]
	\[\mu \defsymbol \{(root,0.05),(0.root,0.1),(1.root,0.1),(2.root,0.1),
		(0.1.root,0.15)\},\]
	\[\phi \defsymbol \{(root,0.5),(0.root,0.7),(1.root,0.65),(2.root,0.6),
		(0.1.root,0.8)\},\]
	\[\rho \defsymbol \{(root,0.5),(0.root,0.65),(1.root,0.5),(2.root,0.6),
		(0.1.root,1)\}.\]
	According to the time functions the configuration $C$ reaches the final
	state in 2.6 time units. The time functions represent the reading, 
	writing, and communication inside and outside containers, respectively.
	\begin{figure}[htbp] 
		\centering
		\begin{subfigure}{.5\textwidth}
			\centering
			\resizebox{4cm}{!}{%
				\begin{tikzpicture}[node distance=2cm, 
				every node/.style={fill=white, font=\sffamily}, align=center]
				\node (root)  [store]                
				{\textit{(root)}\\W $=$ 9};
				\node (0)	  [store, below of=root, left of=root]        
				{\textit{(0)}\\X $\geq$ 11};
				\node (1)  	  [store, below of=root]  
				{\textit{(1)}\\true};
				\node (2)  	  [store, below of=root, right of=root]  
				{\textit{(2)}\\true};
				\node (01)    [store, below of=1] 
				{\textit{(0)}\\Y $>$ 5};
				\draw[->]   (root) -- (0);
				\draw[->]   (root) -- (1);
				\draw[->]   (root) -- (2);
				\draw[->]      (1) -- (01);
				\end{tikzpicture}}
			\caption{Initial state of the system.}
			\label{fig.rtsccs.rtsc-initial}
		\end{subfigure}%
		\begin{subfigure}{.5\textwidth}
			\centering
			\resizebox{4cm}{!}{%
				\begin{tikzpicture}[node distance=2cm,
				every node/.style={fill=white, font=\sffamily}, align=center]
				\node (root)  [store]                
				{\textit{(root)}\\W $=$ 9};
				\node (0)	  [store, below of=root, left of=root]        
				{\textit{(0)}\\X $\geq$ 11};
				\node (1)  	  [store, below of=root]  
				{\textit{(1)}\\true};
				\node (2)  	  [store, below of=root, right of=root]  
				{\textit{(2)}\\Z $\neq$ 10};
				\node (01)    [store, below of=1] 
				{\textit{(0)}\\Y $>$ 5 $\sqcup$\\Y $<$ 10};
				\node (32)    [store, below of=2] 
				{\textit{(3)}\\T $=$ 1};
				\draw[->]   (root) -- (0);
				\draw[->]   (root) -- (1);
				\draw[->]   (root) -- (2);
				\draw[->]      (1) -- (01);
				\draw[->]      (2) -- (32);			
				\end{tikzpicture}}
			\caption{Final state of the system.}
			\label{fig.rtsccs.rtsc-final}
		\end{subfigure}	
		\caption{A containerization example.}
		\label{fig.rtsccs.rtsc}
	\end{figure}
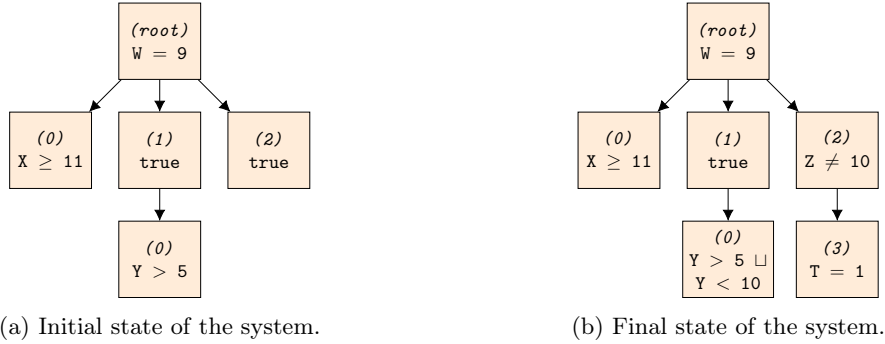  
\end{example}

It is important to note that the timing attributes associated to the
processes in Example~\ref{exa.rtsccp.time} represent constant
time. However, as it will be explained in Section~\ref{sec.rew},
processes can be associated with probability distribution functions to
represent their timing behavior.

\section{Structural Operational Semantics}
\label{sec.sos}

This section introduces the structural operational semantics (SOS) of $\SSCC$.
The SOS shows how the system evolves over time. The inference rules specify
the transitions of the processes and the conditions that must hold for their
execution.


There are some processes in $\SSCC$ that involve consumption of time, such
resource is represented as delay in their execution. This idea comes from
previous works intended to represent timed processes (see --). Processes
including probabilistic choice are supported by an auxiliary function, which
samples a random variable with uniform distribution in order to select,
accordingly to exclusive or independent choice, the processes to be executed.




Time consumption is associated to processes that interact with the store
(querying or adding information) and involve mobility (spatial and extrusion
processes). The time consumed for each process is computed by the function
$\func{\Delta}{Con}{Rat}$, which assigns time to the process accordingly to
the size of the store where it will be executed.


Consider $\Delta(d) = m_d$ for every $d \in Con$. The SOS of $\SSCC$
is defined as follows:


\begin{scriptsize}
\begin{align*}
&
\inference[T-Tell]
{}
{\Conf{\tellp{c}}{d} \to \Conf{\skipp}{\join{d}{\delays{c}{m_d}}}}\\
&
\inference[T-Ask]
{\cleq{c}{d}}
{\Conf{\askp{c}{P}}{d} \to \Conf{\delayp{P}{m_d}}{d}}\\
&
\inference[T-Delay]
{\neg(\cleq{c}{d})}
{\Conf{\askp{c}{P}}{d} \to \Conf{\delayp{\askp{c}{P}}{m_d}}{d}}\\
&\inference[T-Spatial]
{\Conf{P}{d^i} \to \Conf{P'}{d'}}
{\Conf{\K{i}{P}}{d} \to \Conf{\K{i}{\delayp{P'}{m_d}}}{\join{d}{\delays{[d']_i}{m_d}}}}\\
&
\inference[T-Extrusion]
{}
{\Conf{\K{j}{\K{i}{\extr{i}{P}}}} {d} \to \Conf{ \K{j}{\delayp{P}{m_d}}}{d}}\\
&
\inference[Parallel]
{\Conf{P}{d} \to \Conf{P'}{d'}}
{\Conf{\parp{P}{Q}}{d} \to \Conf{\parp{P'}{Q}}{d'}}\\
&
\inference[Exclusive]
{}
{\Conf{\bigoplus_{j \in J}(P_j,q_j)}{d} \to \Conf{P_k}{d}}
\quad
I\left( \bigoplus_{j \in J}(P_j,q_j) \right) = \{k\}\\
&
\inference[Independent]
{}
{\Conf{\bigodot_{j \in J}(P_j,q_j)}{d} \to \Conf{\Pi_{k \in K} P_k}{d}}
\quad
I\left(\bigodot_{j \in J}(P_j,q_j) \right) = K \subseteq J
\end{align*}
\end{scriptsize}


The inference rules of $\SSCC$ represent the execution time of a process as a
delay in its execution. The function $\func{\eta}{Proc \times Rat}{Proc}$ is
used to represent timed processes as delayed processes: $\delayp{P}{m}$ is a
process that mimics $P$ but is delayed $m$ units of time. When $m = 0$, the
process $P$ is launched for execution.
\[
\delayp{P}{m} =
\begin{cases}
P 				& \text{ if } m = 0\\
\delayp{P}{m-t} & \text{ if } m \geq t\\
\end{cases},
\]
In the definition above, $t$ is the execution time of the process with the
least execution time amongst all the process listed for execution.

The function $\func{\eta_s}{Con \times Rat}{Con}$ is defined analogously.
The constraint $\delays{c}{m}$ represents that the constraint $c$ will
not be added to the store until $m$ units of time.
\[
\delays{c}{m} =
\begin{cases}
c 				& \text{ if } m = 0\\
\delays{c}{m-t} & \text{ if } m \geq t\\
\end{cases}
\]

The functions $\eta$ and $\eta_s$ are defined over the rational numbers.
Now, we must verify that $\delayp{P}{0}$ and $\delays{c}{0}$ can be reached,
i.e., they are well-defined.
Recall that $m$ in $\delayp{P}{m}$ and $\delays{c}{m}$ is decreased
$t$ units of time per execution, where $t$ is the execution time of the process with
the least execution time. This time $t$ is extracted from each process in the list of pending processes, then when $P$ is the process with the least time we have $m = t$.
This guarantees that every delayed process will be executed once its 
associated execution time passes.

Notice that delayed constraints are related to {\bf tell} and {\bf spatial}
processes. Both of them reflect a delay to add a constraint to the store.
However, the latter exists because a delayed process within a given space
cannot modifies the local store until its execution time passes.
The function $\func{I}{Proc}{J}$ returns the indexes of the chosen processes
to be executed after the exclusive or independent choice.

The inference rules in the SOS of $\SSCC$ describe the transitions of the
processes. Intuitively, they can be understood as follows (recall that 
$m_d = \Delta(d)$):

\begin{description}
\item[T-Tell] Adds the constraint $c$ to the store through the delayed constraint
$\delays{c}{m_d}$, i.e., adding the constraint is delayed $m_d$ units of time.

\item[T-Ask] If $d$ entails $c$ then it behaves as the process $P$ with a delay of $m_d$ units of time.

\item[T-Delay] If the store $d$ is not strong enough to derive $c$,
the process $\askp{c}{P}$ is delayed $m_d$ units of time.

\item[T-Spatial] If the process $P$ with the information that agent $i$
has in store $d$ (i.e., $d^i$) evolves to $P'$ with store $d'$, then
such a transition can be made within the space of agent $i$ and, the
local store of $i$ is modified with $[d']_i$, both with delay $m_d$.

\item[T-Extrusion] This rule specifies how a process $P$ leaves the space
of agent $i$. It must be known the target space, i.e. $j$'s space.
The delay $m_d$ represents the cost of leaving the space of agent $i$.

\item[Parallel] The process $\parp{P}{Q}$ evolves to $\parp{P'}{Q}$
if a transition can be made from $\Conf{P}{d}$ to $\Conf{P'}{d'}$.

\item[Exclusive] Given a set of processes $\{(P_j,q_j) \mid j \in J\}$,
the function $I$ chooses uniformly an index $k \in J$ of a process
$P_k$ with probability $q_k$. Such process $P_k$ is executed and
the others are precluded.

\item[Independent] Given a set of processes $\{(P_j,q_j) \mid j \in J\}$,
the function $I$ chooses uniformly a subset of indexes $K \subseteq J$
of processes $P_k$ with probability $q_k$ for each $k\in K$.
These processes are executed as the parallel composition
$\Pi_{k \in K} P_k$ and the others are precluded.
\end{description}

As mentioned before, the processes that consume time are those
that model mobility and interact with the store. For that
reason the rules Parallel, Exclusive and Independent are not affected
for the delay $m_d$.

An implementation of the structural operational semantics of $\SSCC$,
using rewriting logic semantics, is presented in the next section.
It provides a formal description of the system language including
its main features of time and probability.

\section{Rewriting Logic Semantics}
\label{sec.rew}

This section presents $\rcal$, a rewriting logic semantics for
$\SSCC$.  Section~\ref{sec.rew.states} includes the key details of the
state infrastructure and Section~\ref{sec.rew.trans} presents the
computational rules of the semantics.  The reader is suggested to
consult on need basis~\ref{sec.rew.scaffolding}, which details aspects
of the constraint system implementation via SMT technology, and
explains the main features of the timing and probabilistic
infrastructure, including the `tick rule' behind the real-time
behavior of the specification. \ref{a.appendix} contains the complete
Maude specification of $\rcal$. This section assumes familiarity with
the Maude language~\cite{clavel-maudebook-2007}.

\subsection{System States}
\label{sec.rew.states}

The rewriting logic semantics for $\PTSCCP$ represents the tree-like
structure of the hierarchical spaces as a flat configuration of
object-like terms. The hierarchical relationships among spaces in
$\PTSCCP$ are captured by using common prefixes as part of an agent's
name. In an observable state, each agent's space is represented by a
set of terms: some encoding the state of execution of all its
processes and exactly one object representing its local store.

The object-based system is represented by the Maude
object~\cite{clavel-maudebook-2007}
predefined by the $\cde{CONFIGURATION}$ module imported in $\cde{SSCC-STATE}$ in
including mode. The objects and configuration are defined using the Maude
object syntax as follows:
\begin{maude}
  subsorts Nat Aid < Oid .
  ops agent process simulation : -> Cid .
\end{maude}
A system state is represented by a configuration of objects containing the 
setup of each one of the agents in the system. A $\cde{Configuration}$ is a 
multiset of objects with set union denoted by juxtaposition and identity 
$\cde{none}$. There are two types of object identifiers (as $\cde{Oid}$): agent 
identifiers (as $\cde{Aid}$) for identifying agents and their hierarchical 
structure, and natural numbers (as $\cde{Nat}$) for some additional 
identification used internally in $\rcal$. There are three types of class 
identifiers (as $\cde{Cid}$), namely, for agents, processes, and a simulation 
object. This object-based representation allows to define a set of attributes, 
which represent the characteristics of the objects as follows:
\begin{maude}
  --- agents attributes
  op store :_    : Boolean          -> Attribute [ctor] .
  op set :_      : Set{Nat}         -> Attribute [ctor] .
  --- processes attributes
  op UID :_      : Nat              -> Attribute [ctor] .
  op command :_  : SSCCCmd          -> Attribute [ctor] .
  --- simulation attributes
  op gtime :_    : Time             -> Attribute [ctor] .
  op pqueue :_   : Heap{2Tuple}     -> Attribute [ctor] .
  op pend :_     : Heap{2Tuple}     -> Attribute [ctor] .
  op nextID :_   : Nat              -> Attribute [ctor] .
  op flag :_     : Bool             -> Attribute [ctor] .
  op counter :_  : Nat              -> Attribute [ctor] .
  op tTM :_      : Map{Aid, StExp}  -> Attribute [ctor] .
  op aTM :_      : Map{Aid, StExp}  -> Attribute [ctor] .
  op sTM :_      : Map{Aid, StExp}  -> Attribute [ctor] .
  op eTM :_      : Map{Aid, StExp}  -> Attribute [ctor] .
  op factor :_   : PosRat           -> Attribute [ctor] .
\end{maude}
Each agent has two attributes, namely, its store (attribute
$\cde{store}$) and a set of its predecessors in the hierarchy
structure (attribute $\cde{set}$); and each process has two
attributes: a universal identifier (used internally for execution
purposes, attribute $\cde{UID}$) and the command (i.e., $\PTSCCP$
process, attribute $\cde{command}$) that it is executing. The
attributes of the simulation object include the global time (attribute
$\cde{gtime}$); the priority queue of system commands to be processed
as ordered by time-to-execution (attribute $\cde{pqueue}$); the
collection of pending commands, i.e., ask commands that are waiting
for its guarding constraint to become active (attribute $\cde{pend}$);
the counter for assigning the next internal identifier when spawning a
new process (attribute $\cde{nextID}$); a flag that is on whenever a
tick rule needs to be applied (attribute $\cde{flag}$); a seed for the
sample of random variables (attribute $\cde{counter}$); a collection
of maps containing the stochastic expression for the time it takes to
process certain commands relative to the space where they are executed
(attributes $\cde{tTM}$, $\cde{aTM}$, $\cde{sTM}$, and $\cde{eTM}$);
and a multiplicative factor for the time it takes to process an ask
command (attribute $\cde{factor}$). The \cde{factor} attribute has
been added to support the fact that querying a store can depend on the
size of its constraint: the bigger it is, the longer it
takes. However, if this feature is not of importance in a specific
application, this attribute can be set to $1$. The sort \cde{Time}, as
it is often the case in Real-time Maude~\cite{olveczky-realtime-2002},
can be used to represent either discrete or dense linear time.

Agent and process objects use a qualified name (sort $\cde{Aid}$) to
identify to which agent's space each one belongs; this sort is defined
in module $\cde{AGENT-ID}$. The hierarchical structure of spaces in
$\SSCC$ is a tree-like structure where the root space is identified by
constant $\cde{root}$. Any other qualified name corresponds to a
dot-separated natural numbers list (sort $\cde{Nat}$), organized from
left to right and including the constant $\cde{root}$ at the end. For
instance, $\cde{3.1.root}$ denotes that agent $\cde{3}$ is child of
agent $\cde{1}$ and, at the same time agent $\cde{1}$ is child of
$\cde{root}$.

\begin{maude}
  op root :         -> Aid .
  op _._  : Nat Aid -> Aid .
\end{maude}

The commands available in the $\PTSCCP$ model are defined in module
$\cde{SSCC-SYNTAX}$, each one of sort $\cde{SSCCCmd}$. Note that the
syntax of each command is very close to the actual syntax in the
$\PTSCCP$ model, e.g., constructs of the form $P \parallel Q$ in $\PTSCCP$ are
represented in the syntax of $\cde{SSCCCmd}$ by terms of the form
$\cde{P || Q}$.

\begin{maude}
  op 0       :                           -> SSCCCmd .
  op tell    : Boolean                   -> SSCCCmd .
  op ask_->_ : Boolean SSCCCmd           -> SSCCCmd .
  op _||_    : SSCCCmd SSCCCmd           -> SSCCCmd [assoc comm gather (e E) ] .
  op _in_    : SSCCCmd Nat               -> SSCCCmd .
  op _out_   : SSCCCmd Nat               -> SSCCCmd .  
  op V       : Nat                       -> SSCCCmd .
  op mu      : Nat     SSCCCmd           -> SSCCCmd .
  op exc     : List{SSCCCmd} List{Float} -> SSCCCmd .
  op ind     : List{SSCCCmd} List{Float} -> SSCCCmd .
\end{maude}
The argument of command $\cde{tell\_}$ is a formula (as
$\cde{Boolean}$), to be added to the agent's constraint. Command
$\cde{ask\_->\_}$ has two arguments: a formula (as $\cde{Boolean}$)
and a program (as $\cde{SSCCCmd}$). The program is executed if the
formula is entailed by the agent's constraint.  Both arguments of
command $\cde{\_||\_}$ are programs (as $\cde{SSCCCmd}$). The
arguments of the command $\cde{\_in\_}$ are a program (as
$\cde{SSCCCmd}$) and a natural number (as $\cde{Nat}$) representing
the identifier of a children (related to an agent $\cde{Aid}$) where
the program is moved. The arguments of the command $\cde{\_out\_}$ are
a program (as $\cde{SSCCCmd}$) and a natural number (as $\cde{Nat}$,
related to an agent $\cde{Aid}$), this command moves a program from an
agent's space to its parent. Command $\cde{V}$ has as argument a
natural number identifying a process local name. The arguments of
command $\cde{mu}$ are a natural number for the variable process to be
replaced, and the program to be replaced in (as $\cde{SSCCCmd}$).
Commands \cde{exc} and \cde{ind} codify exclusive and independent
probabilistic choice; the arguments of the commands $\cde{exc}$ and
$\cde{ind}$ are a list of programs and a list of probabilities (as
$\cde{Float}$ between $0$ and $1$) of the same size, each probability
is related to a program.

\subsubsection{Time functions}
\label{sec.rew.tfuns}

The real-time behavior in $\rcal$ associates timing behavior to those
commands that interact with stores (i.e., $\tell$ and $\ask$ commands)
and to commands that involve mobility among the space structure of the
system (i.e., $\left[\_\right]\_$ and $\extr{\_}{\_}$). More precisely,
$\tell$ and $\ask$ commands take time when posting in and querying
from the store, respectively. Moving the execution of a command inside
an agent and extruding from a space can also take up time, i.e.,
\textit{spatial} and \textit{extrusion} commands can take time. Such
duration is represented by maps from an agent identifier (as
$\cde{Aid}$) to a stochastic expression (as $\cde{StExp}$), i.e., each
agent has its own time functions.  Maps $\cde{tTM}$ (for
\textit{tell}), $\cde{aTM}$ (for \textit{ask}), $\cde{sTM}$ (for
$\left[\_\right]\_$), and $\cde{eTM}$ (for $\extr{\_}{\_}$) can be accessed
using the \cde{getTimeCmd} function. Stochastic expressions include
probability distribution functions to sample the time for the
corresponding command. For example, $\cde{tTM[i]}$ denotes the
stochastic expression to be sampled in order to get the time it takes
to execute a tell command inside the agent's $i$ space.
\begin{maude}
  op fTime : Map{Aid, StExp} Aid Nat -> Tuple{Time, Nat<} .
  eq fTime(TM, L, N)
   = if hasMapping(TM, L) 
     then eval(TM[L], N) 
     else eval(Norm(1.0, 0.2), N) 
     fi .
  op getTimeCmd : SSCCCmd Aid Map{Aid, StExp} Map{Aid, StExp} Map{Aid, StExp} Nat -> Tuple{Time, Nat<} .
  eq getTimeCmd(tell(B1), L, TMt, TMs, TMe, N) = fTime(TMt, L, N) .
  eq getTimeCmd(C1 in I1, L, TMt, TMs, TMe, N) = fTime(TMs, L, N) .
  eq getTimeCmd(C1 out I1, L, TMt, TMs, TMe, N) = fTime(TMe, L, N) .
  eq getTimeCmd(C1, L, TMt, TMs, TMe, N) = (0, N) [owise] .
\end{maude}
The probability distribution functions available for the executable
specification of $\SSCC$ include the exponential, Weibull,
normal/Gauss, $\Gamma$, $\chi^2$, Erlang, F, geometric, Pascal,
Pareto, and uniform functions~\cite{walck-probdist-2007}.

\subsubsection{Probabilistic Choice Functions}
\label{sec.rew.prob}

There are auxiliary functions supporting the probabilistic choice of
processes.

Function \cde{getProb} is used for sampling from the uniform
distribution function for the probabilistic choice.

\begin{maude}
  op getProb : Nat -> Tuple{Float<, Nat<} .
  eq getProb(N) = evalF(Unif(0.0, 1.0), N) .
\end{maude}

The \cde{exclusive} function takes as arguments the list of processes
for the probabilistic choice (as $\cde{List\{SSCCCmd\}}$), the list of
its probabilities (as $\cde{List\{Float\}}$), the counter of the
internal identifier of processes (as $\cde{Nat}$), the priority queue
of the system (as $\cde{Heap\{2Tuple\}}$), the seed for random
sampling (as $\cde{Nat}$), the maps of the time functions (as
$\cde{Map\{Aid, StExp\}}$), and the agent identifier of the agent
where the process is executed (as $\cde{Aid}$). It returns the process
selected to be executed, the next internal identifier to be applied,
the priority queue updated with the selected process, and the new seed
for random sampling in the overall system.

\begin{maude}
  op exclusive : List{SSCCCmd} List{Float} Nat Heap{2Tuple} Nat Map{Aid, StExp} Map{Aid, StExp} Map{Aid, StExp} Aid 
                 -> Tuple{List, Nat, Heap, Nat} .
 ceq exclusive(C, Q, N, P, N1, TMt, TMs, TMe, L) = (C, N + 1, H0, N')
  if (T, N') := getTimeCmd(C, L, TMt, TMs, TMe, N1) /\ H0 := insert(((T, N)), P) .
 ceq exclusive(C NeLC, Q Q1 LF, N, P, N1, TMt, TMs, TMe, L)
   = if Q' <= Q
     then (C, N + 1, H0, N'')
     else exclusive(NeLC, (Q + Q1) LF, N, P, N'', TMt, TMs, TMe, L)
     fi
  if (Q', N') := getProb(N1) /\ (T, N'') := getTimeCmd(C, L, TMt, TMs, TMe, N') /\ H0 := insert(((T, N)), P) .
\end{maude}

Similarly, the \cde{independent} function takes as arguments the list
of processes for the probabilistic choice (as
$\cde{List\{SSCCCmd\}}$), the list of its probabilities (as
$\cde{List\{Float\}}$), the list of previous selected processes (as
$\cde{List\{SSCCCmd\}}$), the counter of the internal identifier of
processes (as $\cde{Nat}$), the priority queue of the system (as
$\cde{Heap\{2Tuple\}}$), the seed for random sampling (as
$\cde{Nat}$), the maps of the time functions (as $\cde{Map\{Aid,
  StExp\}}$), and the agent identifier of the agent where the process
is executed (as $\cde{Aid}$). It returns a list of the processes
selected to be executed, the next internal identifier to be applied,
the priority queue updated with the selected processes, and the new
seed for random sampling.

\begin{maude}
  op independent : List{SSCCCmd} List{Float} List{SSCCCmd} Nat Heap{2Tuple} Nat Map{Aid, StExp} Map{Aid, StExp} 
                   Map{Aid, StExp} Aid -> Tuple{List, Nat, Heap, Nat} .
  eq independent(nil, nil, LC', N, P, N1, TMt, TMs, TMe, L) = (LC', N, P, N1) .
 ceq independent(C LC, Q LF, LC', N, P, N1, TMt, TMs, TMe, L)
   = if Q' <= Q
     then independent(LC, LF, LC' C, N + 1, H0, N'', TMt, TMs, TMe, L)
     else independent(LC, LF, LC', N, P, N'', TMt, TMs, TMe, L)
     fi
  if (Q', N') := getProb(N1) /\ (T, N'') := getTimeCmd(C, L, TMt, TMs, TMe, N') /\ H0 := insert(((T, N)), P) .
\end{maude}

\subsection{System Transitions}
\label{sec.rew.trans}

The transitions in the rewriting logic semantics of $\SSCC$ comprise
both \emph{invisible} (given by equations) and \emph{observable}
transitions (given by rules).

There are two invisible transitions, namely, one for removing a
$\cde{0}$ command from a configuration and another one to join the
contents of two stores of the same space (i.e., two stores with the
same $\cde{Aid}$).

\begin{maude}
  eq < L0 : process | command : 0, Atts > = none .
  eq < L0 : agent | store : B0 > < L0 : agent | store : B1 > = < L0 : agent | store : (B0 and B1) > .
\end{maude}

The second type of (invisible) transitions is important because when a
new process is spawned in an agent's space, a store with the empty
constraint (i.e., $\cde{true}$) is created for that space. If such a
space existed before, then the idea is that the newly created store is
subsumed by the existing one. Note that neither of the invisible
transitions take time, i.e., they are instantaneous and axiomatize
structural properties of commands.

There are nine observable transitions, i.e., rewrite rules, that
capture the concurrent behavior in $\rcal$; they are explained in the
rest of this section. It is important to observe that rules such as
\cde{[exclusive]} and \cde{[independent]} are probabilistic in
nature. However, the syntax of probabilistic rewrite rules is slightly
modified with respect to the one presented in Section~\ref{sec.prelim}
by encapsulating probabilistic sampling is some auxiliary functions
presented before. Furhtermore, the stochastic behavior of processes is
determined also by auxiliary functions that sample probability
distribution functions to assign a time (i.e., duration) to processes.
Finally, in most of the rules, the counter in the simulation object is
updated with a new value computed in the conditions. This is because
more than one probabilistic choice could be performed internally by
the auxiliary functions in the conditions; in these cases, the new
counter represents the next counter that can be used after all of
these samplings have been made.

\paragraph{Rule \cde{[tell]}} It defines the semantics of $\tell(\_)$
processes. Once a process of this type is the next to be executed (as
indicated by the priority queue in the \cde{simulation} object with
time $0$), its constraint is placed in the corresponding store and
the process terminates.

\begin{maude}
  rl [tell] :
     < L0 : agent | store : B0 >
     < L0 : process | UID : I0, command : tell (B1) >
     < I : simulation | pqueue : T(Ra,((Ti, I0)), Le, Ri), flag : false, pend : P, Atts >
  => < L0 : agent | store : (B0 and B1) >
     < I : simulation | pqueue : T(Ra,((Ti, I0)), Le, Ri), flag : true, pend : P, Atts > .
\end{maude}

\paragraph{Rule \cde{[ask]}}
It defines the semantics of $\ask(\_) \rightarrow \_$ commands when
their guards are entailed by the constraint in the current store.
Note that the semantic consequence relation of the constraint system
is queried by asking the SMT solver. When the guard $\cde{B1}$ is
entailed by the constraint $\cde{B0}$ in the local store $\cde{L0}$,
the command $\cde{C1}$ is moved into the priority queue of the system
(attribute $\cde{pqueue}$).  The time of the command $\cde{C1}$
includes the time function $\cde{aTM}$ for querying the local store
$\cde{B0}$ and a factor of its size (attribute $\cde{factor}$). Note
that the matching conditions (i.e., the ones of the form \cde{\_ :=
  \_}) are used in a similar way as ``let'' commands in functional
programming languages.

\begin{maude}
 crl [ask] :
     < L0 : agent | store : B0 >
     < L0 : process | UID : I0, command : (ask B1 -> C1) >
     < I : simulation | pqueue : T(Ra,((Ti, I0)), Le, Ri), flag : false, pend : P, nextID : N, counter : N1, 
                        tTM : TMt, aTM : TMa, sTM : TMs, eTM : TMe, factor : alpha, Atts >
  => < L0 : agent | store : B0 >
     < L0 : process | UID : N, command : C1 >
     < I : simulation | pqueue : T(Ra,((Ti, I0)), Le, Ri), flag : true, pend : H0, nextID : (N + 1), counter : N3, 
                        tTM : TMt, aTM : TMa, sTM : TMs, eTM : TMe, factor : alpha, Atts >
  if entails(B0,B1) /\ (T0, N2) := getTimeCmd(C1, L0, TMt, TMs, TMe, N1) /\ 
     (T1, N3) := fTime(TMa, L0, N2) /\ S := size(B0) /\ H0 := insert(((T0 plus (T1 plus (S * alpha)), N)), P) .
\end{maude}

\paragraph{Rule \cde{[delay]}}
It defines the semantics of $\ask(\_) \rightarrow \_$ commands when
their guards are \textit{not} entailed by the constraint in the
current store.  Similar to the case handled by the rule \cde{[ask]},
the semantic consequence relation of the constraint system is queried
by asking the SMT solver. When the guard $\cde{B1}$ is not entailed by
the constraint $\cde{B0}$ in the local store $\cde{L0}$, the ask
command is delayed and placed in the priority queue for pending ask
commands, where it will remain ``locked'' until the $\rlname{tick}$
rule executes again (see~\ref{sec.rew.tick}).

\begin{maude}
 crl [delay] :
     < L0 : agent | store : B0 > < L0 : process | UID : I0, command : (ask B1 -> C1) >
     < I : simulation | pqueue : T(Ra,((Ti, I0)), Le, Ri), pend : P, Atts >
  => < L0 : agent | store : B0 > < L0 : process | UID : I0, command : (ask B1 -> C1) >
     < I : simulation | pqueue : merge(Le, Ri), pend : insert(((Ti, I0)),P), Atts >
  if not(entails(B0,B1)) .
\end{maude}

\paragraph{Rule \cde{[parallel]}}
It implements the semantics for parallel composition of processes by
spawning the two processes in the current space by creating a new
object in the configuration for each of the two commands. These two
commands are assigned a time-to-execution and added to the system's
scheduler.

\begin{maude}
 crl [parallel] :
     < L0 : process | UID : I0, command : (C0 || C1) >
     < I : simulation | pqueue : T(Ra,((Ti, I0)), Le, Ri), nextID : N, flag : false, pend : P, counter : N1, 
                        tTM : TMt, sTM : TMs, eTM : TMe, Atts >
  => < L0 : process | UID : N, command : C0 >
     < L0 : process | UID : (N + 1), command : C1 >
     < I : simulation | pqueue : T(Ra,((Ti, I0)), Le, Ri), nextID : (N + 2), flag : true, pend : H0, counter : N3, 
                        tTM : TMt, sTM : TMs, eTM : TMe, Atts >
  if (T0, N2) := getTimeCmd(C0, L0, TMt, TMs, TMe, N1) /\ (T1, N3) := getTimeCmd(C1, L0, TMt, TMs, TMe, N2) /\ 
     H0 := insert(((T0, N)), insert(((T1, N + 1)), P)) .
\end{maude}

\paragraph{Rule \cde{[recursion]}}
It defines the semantics of the $\mu \_.\_$ recursion
commands. Operationally, for a command \cde{mu(N0,C0)} ready for
execution, it replaces all appearances of the variable command
$\cde{V(N0)}$ within command $\cde{C0}$ with $\cde{mu(N0, C0)}$ using
the auxiliary function $\cde{replace}$. This creates the effect of a
recursive call.

\begin{maude}
 crl [recursion] :
     < L0 : process | UID : I0, command : mu(N0, C0) >
     < I : simulation | pqueue : T(Ra,((Ti, I0)), Le, Ri), nextID : N, flag : false, pend : P, counter : N1, 
                        tTM : TMt, sTM : TMs, eTM : TMe, Atts >
  => < L0 : process | UID : N, command : replace(N0, C0, mu(N0,C0)) >
     < I : simulation | pqueue : T(Ra,((Ti, I0)), Le, Ri), nextID : (N + 1), flag : true, pend : H0, counter : N2, 
                        tTM : TMt, sTM : TMs, eTM : TMe, Atts >
  if (T0, N2) := getTimeCmd(C0, L0, TMt, TMs, TMe, N1) /\ H0 := insert(((T0, N)), P) .
\end{maude}

\paragraph{Rules \cde{[extrusion]} and \cde{[space]}}
They define the semantics of space navigation given by commands
$\extr{\_}{\_}$ and $\K{\_}{\_}$. The $\rlname{extrusion}$ rule
executes the extrusion command \cde{CO out NO} by executing the
$\cde{C0}$ command in the parent's space $\cde{L0}$ of the current
space $\cde{N0}$. The \cde{[space]} rule executes the command \cde{C0
  in N0} by creating a new space for agent \cde{N0} inside the current
space (denoted by $\cde{N0.L0}$) with an empty store (i.e.,
$\cde{true}$), in addition to the execution of the $\cde{C0}$ command
within the new agent's space. Note that if such a space already
exists, the empty store will be subsumed by it thanks to the invisible
rule for merging two stores corresponding to the same agent, as
presented above. If the time functions are not defined for the new
agent in the initial state, then they are inherited from its nearest
ancestor for future computation; otherwise, they are assigned to a
default distribution function.

\begin{maude}     
 crl [extrusion]:
     < N0 . L0 : process | UID : I0, command : (C0 out N0) >
     < I : simulation | pqueue : T(Ra,((Ti, I0)), Le, Ri), nextID : N, flag : false, pend : P, counter : N1, 
                        tTM : TMt, sTM : TMs, eTM : TMe, Atts >
  => < L0 : process | UID : N, command : C0 >
     < I : simulation | pqueue : T(Ra,((Ti, I0)), Le, Ri), flag : true, pend : H0, nextID : (N + 1), counter : N2, 
                        tTM : TMt, sTM : TMs, eTM : TMe, Atts >
  if (T0, N2) := getTimeCmd(C0, L0, TMt, TMs, TMe, N1) /\ H0 := insert(((T0, N)), P) .

 crl [space] :
     < L0 : process | UID : I0, command : (C0 in N0) >
     < I : simulation | pqueue : T(Ra,((Ti, I0)), Le, Ri), nextID : N, flag : false, pend : P, counter : N1, 
                        tTM : TMt, aTM : TMa, sTM : TMs, eTM : TMe, Atts >
  => < N0 . L0 : agent | store : true >
     < N0 . L0 : process | UID : N, command : C0 >
     < I : simulation | pqueue : T(Ra,((Ti, I0)), Le, Ri), flag : true, pend : H0, nextID : (N + 1), counter : N2, 
                        tTM : TMt', aTM : TMa', sTM : TMs', eTM : TMe', Atts >
  if TMt' := insert(N0 . L0, get-ancestor(TMt, N0 . L0), TMt) /\ 
     TMa' := insert(N0 . L0, get-ancestor(TMa, N0 . L0), TMa) /\
     TMs' := insert(N0 . L0, get-ancestor(TMs, N0 . L0), TMs) /\ 
     TMe' := insert(N0 . L0, get-ancestor(TMe, N0 . L0), TMe) /\
     (T0, N2) := getTimeCmd(C0, N0 . L0, TMt', TMs', TMe', N1) /\ 
     H0 := insert(((T0, N)), P) .
\end{maude}

\paragraph{Rule \cde{[exclusive]}}
It implements the semantics of the $\bigoplus\_$ exclusive
probabilistic choice command. This rule executes a single command from
a given list of commands $\cde{LC}$ with their corresponding
probabilities $\cde{LF}$. For each command in the list, a random
variable with uniform distribution is sampled by calling the auxiliary
function \cde{exclusive}. Internally, if the sampled probability is
greater or equal to the probability associated to the corresponding
command, then it is selected for execution; otherwise, the next
command is evaluated with the corresponding cumulative probability. In
this function, it is assumed that the sum of the probabilities of the
list must be equal to 1. The last command of the list is executed if
no other one is executed.

\begin{maude}     
 crl [exclusive] :
     < L0 : process | UID : I0, command : exc(LC, LF) >
     < I : simulation | pqueue : T(Ra,((Ti, I0)), Le, Ri), nextID : N, flag : false, pend : P, counter : N1, 
                        tTM : TMt, sTM : TMs, eTM : TMe, Atts >
  => genCommands(LC', N, L0) 
     < I : simulation | pqueue : T(Ra,((Ti, I0)), Le, Ri), nextID : N', flag : true, pend : P0, counter : N'', 
                        tTM : TMt, sTM : TMs, eTM : TMe, Atts >
  if (LC', N', P0, N'') := exclusive(LC, LF, N, P, N1, TMt, TMs, TMe, L0) .
\end{maude}

\paragraph{Rule \cde{[independent]}}
It implements the semantics of the $\bigodot\_$ independent
probabilistic choice command. This rule executes a subset of a given
list of commands $\cde{LC}$ with their corresponding probabilities
$\cde{LF}$ with the help of the auxiliary function
\cde{independent}. Each command in the list is chosen or dropped by
using a random variable with uniform distribution. In the end, all,
some, or none of the given commands in the list can be selected for
execution.

\begin{maude}
 crl [independent] :
     < L0 : process | UID : I0, command : ind(LC, LF) >
     < I : simulation | pqueue : T(Ra,((Ti, I0)), Le, Ri), nextID : N, flag : false, pend : P, counter : N1, 
                        tTM : TMt, sTM : TMs, eTM : TMe, Atts >
  => genCommands(LC', N, L0) 
     < I : simulation | pqueue : T(Ra,((Ti, I0)), Le, Ri), nextID : N', flag : true, pend : P0, counter : N'', 
                        tTM : TMt, sTM : TMs, eTM : TMe, Atts >
  if (LC', N', P0, N'') := independent(LC, LF, nil, N, P, N1, TMt, TMs, TMe, L0) .
\end{maude}

This section concludes by introducing an example illustrating the
rewriting logic semantics developed for $\SSCC$.

\begin{example}\label{exa.rew.case}
  The initial state $(a)$ in Figure~\ref{fig.rtsccs.rtsc} can
  be represented in the $\SSCC$ semantics as follows:
	\begin{maude}
  < root : agent | store : W === 9 >
  < 0 . root : agent | store : X >= 11 >
  < 1 . root : agent | store : true >
  < 2 . root : agent | store : true >
  < 0 . 1 . root : agent | store : Y > 5 >
	\end{maude}
  In this syntax, the hierarchical task assignment system is modeled,
  where agents are workers and processes are tasks to be
  assigned. Mobility in this system is modeled by tasks flowing
  between workers.  Each worker has a unique representation with a
  unique qualified name (i.e., \cde{Aid}). This representation is
  useful to identify the time taken by each one of the
  processes. Consider the following process $R$:
  \begin{itemize}[noitemsep]
    \scriptsize
    \item [$P_{1_1} :=$] \texttt{ind((tell(A === 1) in 1) (tell(B === 1) in 2) (tell(C === 1) in 3) (tell(D === 1) in 4), 0.5 0.5 0.5 0.5)}
    \item [$P_{1_2} :=$] \texttt{(tell(Y === 25) || ask(Y > 2) -> (tell(Y > 2) out 2)) in 2}
    \item [$P_1 :=$] \texttt{exc(((}$P_{1_1}$ \texttt{|| tell(Y === 5) || (ask(Y > 2) -> (tell(Y > 2) out 1))) in 1) } $P_{1_2}$ \texttt{, 0.60 0.40)}
    \item [$P_2 :=$] \texttt{ask (Y > 2) -> (tell(X === 15) || ask(X >= 10) -> (tell(X >= 10) out 1)}
    \item [$P :=$] \texttt{(}$P_1$ \texttt{||} $P_2$\texttt{) in 1}
    \item [$Q_1 :=$] \texttt{(tell(Z === 9) || ask(Z < 15) -> (tell(Z < 15) out 3)) in 3}
    \item [$Q_2 :=$] \texttt{(tell(W === 25) || ask(W > 0) -> (tell(W > 0) out 4)) in 4}
    \item [$Q_3 :=$] \texttt{ask (Z < 15 and W > 0) -> (tell(V === 67) || ask(V < 100) -> (tell(V < 100) out 2)}
    \item [$Q :=$] \texttt{(}$Q_1$ \texttt{||} $Q_2$ \texttt{||} $Q_3$\texttt{) in 2}
    \item [$R :=$] $P$ \texttt{||} $Q$ \texttt{|| (ask (X >= 10 and V < 100) -> (tell(U === 50) || ask(U < 55) -> tell(DONE)))}
  \end{itemize}  
  The hierarchical system for process $R$ is represented in
  Figure~\ref{fig.rls-ex}. This configuration is encoded as follows:
  \begin{maude}
  < root : agent | store : true >
  < root : process | UID : 1, command : R >
  < 1 : simulation | gtime : 0,pqueue : T(1,((0,1)),empty,empty),pend : empty,nextID : 19, flag : false, 
                     counter : N, tTM : ((root) |-> Norm(1.0, 0.2)), aTM : ((root) |-> Norm(1.2, 0.2)) , 
                     sTM : ((root) |-> Norm(0.5, 0.2)) , eTM : ((root) |-> Norm(0.5, 0.2)), factor : 1/2 >
  \end{maude}

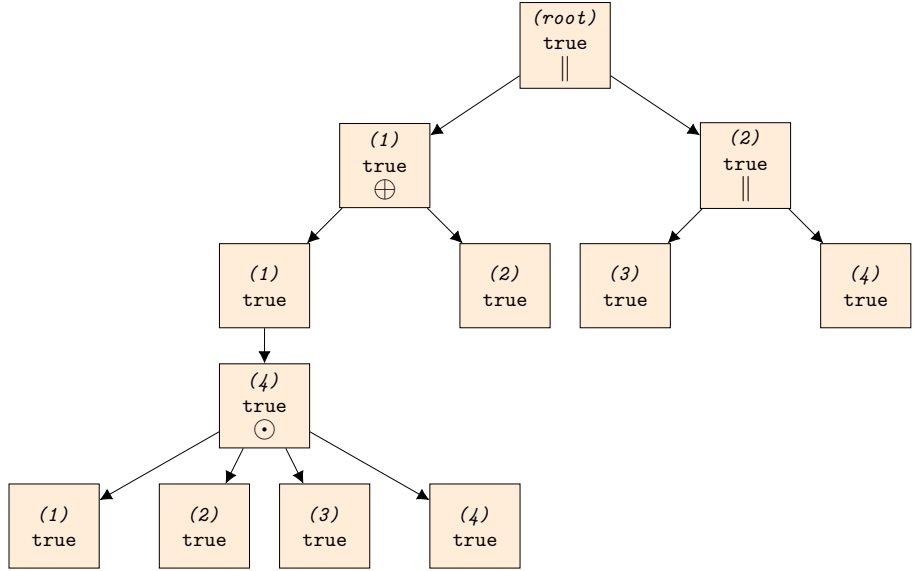
\begin{figure}[htbp] 
  \centering
  \resizebox{12cm}{!}{%
    \begin{tikzpicture}[node distance=2cm, 
    every node/.style={fill=white, font=\sffamily}, align=center]
    \node (root)  [store]
    {\textit{(root)}\\true\\{\large $\parallel$}};
    \node (1)	  [store, below of=root, left of=root, xshift=-1cm]
    {\textit{(1)}\\true\\{\Large $\oplus$}};    
    \node (2)  	  [store, below of=root, right of=root, xshift=1cm]  
    {\textit{(2)}\\true\\{\large $\parallel$}};
    \node (23)	  [store, below of=2, left of=2]
    {\textit{(3)}\\true};
    \node (24)  	  [store, below of=2, right of=2]
    {\textit{(4)}\\true};
    \node (11)	  [store, below of=1, left of=1]
    {\textit{(1)}\\true};
    \node (12)  	  [store, below of=1, right of=1]
    {\textit{(2)}\\true};
    \node (114)	  [store, below of=11]
    {\textit{(4)}\\true\\{\Large $\odot$}};
    \node (1141)	  [store, below of=114, left of=114, xshift=-1.5cm]
    {\textit{(1)}\\true};
    \node (1142)	  [store, below of=114, left of=114, xshift=1cm]
    {\textit{(2)}\\true};
    \node (1143)	  [store, below of=114, right of=114, xshift=-1cm]
    {\textit{(3)}\\true};
    \node (1144)	  [store, below of=114, right of=114, xshift=1.5cm]
    {\textit{(4)}\\true};    
    \draw[->]  (root) -- (1);
    \draw[->]  (root) -- (2);
    \draw[->]     (1) -- (11);
    \draw[->]     (1) -- (12);
    \draw[->]     (2) -- (23);
    \draw[->]     (2) -- (24);
    \draw[->]    (11) -- (114);
    \draw[->]   (114) -- (1144);
    \draw[->]   (114) -- (1143);
    \draw[->]   (114) -- (1142);
    \draw[->]   (114) -- (1141);
    \end{tikzpicture}}
  \caption{Example. Initial state of the system.}
  \label{fig.rls-ex}
\end{figure}
	The global time of the system (attribute $\cde{gtime}$) starts at
  zero. The random seed for the probability distributions sampling
  (attribute $\cde{counter}$) is $\cde{N}$ and the multiplicative
  factor (attribute $\cde{factor}$) is $\cde{1/2}$, i.e., the
  $\rlname{ask}$ rule will consider half of the size of the local
  store of the agent. The size of an agent's store is proportional to
  the number of atomic quantifier-free formulas in it plus its Boolean
  connectives. For example, the size of the formula $\cde{W > 0 and Z
    < 15}$, with \texttt{W} and \texttt{Z} integers, is $3$, therefore
  the $\rlname{ask}$ includes $1.5$ time units. Time functions
  ($\alpha, \mu, \phi, \rho$) are defined as follows, respectively:
	\begin{maude}
  tTM : ((root) |-> Norm(1.0, 0.2)), 
  aTM : ((root) |-> Norm(1.2, 0.2)),
  sTM : ((root) |-> Norm(0.5, 0.2)), 
  eTM : ((root) |-> Norm(0.5, 0.2))
	\end{maude}
  The time functions of the example are normal distributions with the
  usual median and variance parameters, respectively. This expression
  is defined for the root agent and the other agents will inherit
  it. For example, querying the store at the root agent uses a normal
  distribution function with parameters $\mu=1.2$ and $\sigma^2=0.2$.

	The system may evolve as detailed below. Note that the final state
  depends on the random sampling of the time functions and the
  $\rlname{exclusive}$ and $\rlname{independent}$ rules.
	\begin{maude}
  < root : agent | store : (X:Integer >= 10 and V:Integer < 100 and U:Integer === (50).Integer and DONE:Boolean) >
  < 1 . root : agent | store : (Y:Integer > 2 and X:Integer === (15).Integer) >
  < 2 . root : agent | store : (W:Integer > 0 and Z:Integer < 15 and V:Integer === (67).Integer) > 
  < 1 . 1 . root : agent | store : (Y:Integer === (5).Integer) > 
  < 4 . 2 . root : agent | store : (W:Integer === (25).Integer) >   
  < 3 . 2 . root : agent | store : (Z:Integer === (9).Integer) > 
  < 4 . 1 . 1 . root : agent | store : true > 
  < 2 . 4 . 1 . 1 . root : agent | store : (B:Integer === (1).Integer) > 
  < 3 . 4 . 1 . 1 . root : agent | store : (C:Integer === (1).Integer) >   
  < 4 . 4 . 1 . 1 . root : agent | store : (D:Integer === (1).Integer) >
	\end{maude}
  Figure~\ref{fig.rls-ex} depicts the exclusive process $P_1$ in the
  space of agent $\cde{1.root}$. In the final state encoded above, the
  agent $\cde{1.1.root}$ was created as a result of executing the
  $P_1$ process. Agents $\cde{2.4.1.1.root}$, $\cde{3.4.1.1.root}$,
  and $\cde{4.4.1.1.root}$ were independently from process $P_{1_1}$.
\end{example}

\section{Model Simulation}
\label{sec.reach}

Some properties of reactive systems, e.g., fault-tolerance and
consistency, can be tested with the rewriting logic semantics $\rcal$
of $\SSCC$. In particular, this section uses Maude's \texttt{search}
command to test for consistency, fault-tolerance, and knowledge
inference examples. Note, however, that since the rewrite theory
$\rcal$ is probabilistic in nature, the results in this section can
change if the seed selected for executing the experiments is
different. The reader is referred to Section~\ref{sec.case} for an
example of quantitative analysis with $\rcal$ in the form of
statistical model checking.

This section uses a $\SSCC$ simplified version of the
system presented in Example~\ref{exa.rew.case}, where:

\begin{itemize}[noitemsep]
  \scriptsize
  \item [$P_{1_1} :=$] \texttt{ask (W > 1) -> (tell(Y === 32) || ask(Y > 9) -> 
    (tell(Y > 9) out 2))}
  \item [$P_1 :=$] \texttt{(tell (W ===5) \texttt{||} $P_{1_1}$) in 2}
  \item [$P_2 :=$] \texttt{ask (Y > 2) -> (tell(X === 15) || ask(X >= 10) -> 
  (tell(X >= 10) out 1))}
  \item [$P :=$] \texttt{(}$P_1$ \texttt{||} $P_2$\texttt{) in 1}
  \item [$Q :=$] \texttt{ask (X >= 10) -> (tell(U === 50) 
    || ask(U < 55) -> tell(DONE))}
  \item [$R :=$] $P$ \texttt{|| $Q$}
\end{itemize}  

The process $P_1$ posts some information in the space of agent
$\cde{2.1.root}$ and waits until it has enough information to infer $W
> 1$. Once it has enough information, it posts $Y > 9$ in the space of
its ancestor. Process $P$ executes $P_1$ and asks if $Y > 2$ is known
to be true in the space of agent $\cde{1.root}$. Once the agent has
gained enough information, it posts $X\geq 10$ in the space of its
ancestor.  Process $R$ executes $P$ and asks if $X \geq 10$ is known
to be true in the space of agent $\cde{root}$. Once this agent has
gained enough information, it posts $\texttt{DONE}$ in its current
space. Note that probabilities are used only for sampling the
execution time of processes, and that exclusive and independent
operators are not used.


\paragraph{Consistency and Fault-tolerance}

Consistency is the property that ensures a local failure does not propagate to 
the entire system; fault tolerance ensures a system to continue operating 
properly in the event of a failure. In $\SSCC$, this means that if a store 
becomes inconsistent, it is not the case that such an inconsistency spreads to 
the entire system. Even though, inconsistencies can appear in other stores due 
to some unrelated reasons. 


Queries such as these ones, can be implemented with the help of
$\rcal$ and the rewriting modulo SMT approach by using Maude's
\cde{search} command.  As an example, consider the following
\cde{search} command:
\begin{maude}
  search in APMAUDE :
    < root : agent | store : (X < 5), set : (empty) >
    < root : process | UID : 1, command : R >
    < 1 : simulation | gtime : 0,pqueue : T(1,((0,1)),empty,empty),pend : empty,nextID : 2, flag : false, 
                       counter : 13, tTM : ((root) |-> Norm(1.0, 0.2)), aTM : ((root) |-> Norm(1.2, 0.2)) ,
                       sTM : ((root) |-> Norm(0.5, 0.2)) , eTM : ((root) |-> Norm(0.5, 0.2)),factor : 1/2 >
    flg(true, 1000.0)
    =>* < L0 : agent | store : B0:Boolean, set : SN:Set{Nat} > C:Config
    such that check-unsat(B0:Boolean) .
\end{maude}
\noindent Note that a store is inconsistent if it is unsatisfiable,
thereby checking whether a store is inconsistent is accomplished with
the function \cde{check-unsat}. This command finds 20 reachable states
where there is at least one inconsistent store. Therefore, even though
the inconsistency appears, the system continues evolving until no more
processes can be performed. It is possible to verify that there are
states with consistent and inconsistent stores at the same time by
slightly modifying the search command.

\begin{maude}
  Solution 1 (state 341)
  states: 342  rewrites: 276047 in 1888ms cpu (1889ms real) (146211 rewrites/second)
  C:Config --> flg(true, 1.0e+3) 
    < root : process | UID : 3,command : Q > 
    < 1 : simulation | gtime : 16.22,pqueue : T(1,(0.98,21),empty,empty),pend : T(1,(0,3),empty,empty),
                       nextID : 22,counter : 26,flag : true,tTM : ...,aTM : ...,sTM : ...,eTM : ...,factor : 1/2 > 
    < 1 . root : agent | store : (Y:Integer > 9 and X:Integer === (15).Integer),set : 2 > 
    < 2 . 1 . root : agent | store : (W:Integer === (5).Integer and
  Y:Integer === (32).Integer),set : empty >
  L0 --> root
  B0 --> X:Integer < (5).Integer and X:Integer >= (10).Integer
  SN:Set{Nat} --> (1).NzNat
  ...
\end{maude}

\paragraph{Knowledge Inference}
It refers to acquiring new knowledge from existing facts. In the
setting of $\rcal$, this means at some point an agent has gained
enough information to infer --from the rules of first-order logic--
new facts.
As an example, consider the following \cde{search} command:

\begin{maude}
  search in APMAUDE :
    < root : agent | store : true, set : (empty) >
    < root : process | UID : 1, command : R >
    < 1 : simulation | gtime : 0,pqueue : T(1,((0,1)),empty,empty),pend : empty,nextID : 2, flag : false, 
                       counter : 13, tTM : ((root) |-> Norm(1.0, 0.2)), aTM : ((root) |-> Norm(1.2, 0.2)) ,
                       sTM : ((root) |-> Norm(0.5, 0.2)) , eTM : ((root) |-> Norm(0.5, 0.2)), factor : 1/2 >
    flg(true, 1000.0)
  =>* < L0 : agent | store : B0:Boolean, set : SN:Set{Nat} > C:Config
  such that entails(B0:Boolean, gen-var("X") > 9) .
\end{maude}

\noindent This command finds a reachable state from the given initial
state, in which some store logically implies $Y > 9$. This query finds
the following solutions:

\begin{maude}
  Solution 1 (state 211)
  states: 212  rewrites: 241358 in 1156ms cpu (1160ms real) (208787 rewrites/second)
  C:Config --> flg(true, 1.0e+3) 
    < root : agent | store : true,set : 1 > 
    < root : process | UID : 3,command : Q > 
    < 1 : simulation | gtime : 4.41,pqueue : T(1,(1.03,14),empty,empty),
                       pend : T(2,(0,3),T(1,(0,13),empty,empty),T(1,(0,7),empty,empty)),nextID : 15,
                       counter : 18,flag : true,tTM : ...,aTM : ...,sTM : ...,eTM : ..,factor : 1/2 > 
    < 1 . root : agent | store : true, set : 2 > 
    < 1 . root : process | UID : 7,command : P2 > 
    < 2 . 1 . root : process | UID : 13,command : (ask Y:Integer > 9 -> (tell(Y:Integer > (9).Integer) out 2)) >
  L0 --> 2 . 1 . root
  B0 --> W:Integer === (5).Integer and Y:Integer === (32).Integer
  SN:Set{Nat} --> (empty).Set{Nat}
  ...
\end{maude}

\paragraph{Same Knowledge}
Same knowledge refers to different agents gaining, at some point, the
same knowledge. In the case of the particular constraint system
implemented in this manuscript, it means having two stores with
logically equivalent stores.
As an example, consider the following Maude \cde{search} command,
querying for two stores having the same information when they are
non-empty:

\begin{maude}
  search in APMAUDE :
    < root : agent | store : true, set : (empty) >
    < root : process | UID : 1, command : R >
    < 1 : simulation | gtime : 0,pqueue : T(1,((0,1)),empty,empty),pend : empty,nextID : 2, flag : false, 
                       counter : 13,tTM : ((root) |-> Norm(1.0, 0.2)), aTM : ((root) |-> Norm(1.2, 0.2)) ,
                       sTM : ((root) |-> Norm(0.5, 0.2)) , eTM : ((root) |-> Norm(0.5, 0.2)), factor : 1/2 >
    flg(true, 1000.0)
  =>* < L0 : agent | store : B0:Boolean, set : SN:Set{Nat} > 
      < L1 : agent | store : B1:Boolean, set : SN1:Set{Nat} > C:Config
  such that entails(B0:Boolean, B1:Boolean) /\ entails(B1:Boolean, B0:Boolean) /\ B1:Boolean =/= true .
\end{maude}
\noindent Note that it is never the case that there are two stores with the same
information, which agrees with the following output of Maude.

\begin{maude}
  No solution.
  states: 363  rewrites: 306427 in 4824ms cpu (4859ms real) (63521 rewrites/second)
\end{maude}

It is important to emphasize, as pointed out in the opening of this
section, that the fact that no solution can be found, does not mean
that actually no solution exists: this reachability queries are
dependent on the seed used for the pseudo-random number
generator. That is, the reachability queries presented in this section
can be seen as mechanisms for testing, but not as a fully-fledged
approached for reachability analysis.

\section{Statistical Model Checking of a Random Search on a Hierarchy of Spaces: A Case Study}
\label{sec.case}

This section presents a case study on a process that performs a random
search on a hierarchy of spaces, which illustrates both the use of the
$\SSCC$ rewriting logic semantics and how statistical model checking
can be performed with the help of the $\PVESTA$ statistical model
checker.

\subsection{Description of the Case Study}

Social networks relate agents sharing information with each
other. Such interactions could create, e.g., friend circles, social
forums, and debates. A discussion among agents could result in a
far-from-desirable-scenario in which damaging comments are
posted. This could ultimately affect users in the social network.
Therefore, the possibility to detect such unwelcome posts can be seen
as a mechanism to improve fair interaction among users in the social
network.

The case study is about a robot trying to find inappropriate posts in
a social network, by randomly visiting the spaces associated to each
user. As such, the social network is represented in the $\SSCC$ model
as a hierarchy of spaces and the information posted by the users,
distributed in the hierarchy of spaces, as constraints.  The overall
idea is that the robot will explore the network looking for posts
containing unwanted information. The evolution of the robot in the
space hierarchy is decided at random by visiting the spaces adjacent
to the one it is with equal probability (i.e., it chooses from the
parent space --if it exists-- and the children spaces uniformly at
random). If such a post is found, a constraint is added to the
corresponding store representing the fact that a warning message is
posted. To keep the case study simple, the robot will stop once it
finds the first unwanted post. If more posts were to be found, a copy
of the robot process could be spawn again.

\subsection{Formal Specification}

The behavior of the robot is specified by two rewrite rules. Each one
of these rules implements a macro command called \cde{watch} that is
defined on top of the $\SSCC$ specification and, as such, does not add
any new expressive power to the model. Specifically, the command
\cde{watch(C, B)} indicates that if the store in which the constraint
$\cde{B}$ is entailed (e.g., an offensive message is found), the
command $\cde{C}$ is the action to be executed by the robot.

The \cde{[search]} rule handles the \cde{watch} command when the
undesired post is not found in the current space.

\begin{maude}
	crl [search] :
	  < L0 : agent | store : B0, set : SN >
	  < L0 : process | UID : I0, command : watch(C0, B1) >
	  < I : simulation | pqueue : T(Ra,((Ti, I0)), Le, Ri), nextID : N,
	    flag : false, pend : P, counter : N1, tTM : TMt, sTM : TMs, eTM : TMe, Atts >
	=>
	  < L0 : agent | store : B0, set : SN >
	  < L0 : process | UID : N, command : exc(LC, LF) >
	  < I : simulation | pqueue : T(Ra,((Ti, I0)), Le, Ri), nextID : (N + 1),
	    flag : true, pend : H0, counter : N'', tTM : TMt, sTM : TMs, eTM : TMe, Atts >
	 if not(entails(B0,B1))
	 /\ LC := command-list(L0, SN, watch(C0, B1))
	 /\ P(LF, N') := prob-list(size(LC), N1)
	 /\ (T0, N'') := getTimeCmd(exc(LC, LF), L0, TMt, TMs, TMe, N')
	 /\ H0 := insert(((T0, N)), P) .
\end{maude}	

\noindent The \cde{[search]} rule interprets the command
$\cde{watch(C0, B1)}$ into an exclusive command $\cde{exc(LC, LF)}$
when the constraint $\cde{B1}$ is not entailed by the store in the
space where the robot is. The exclusive command uniformly at random
chooses the next space the robot will visit, by either leaving the
current space or going to a children space. If no such a space exists,
then it will stay in its current space.

The rule $\cde{[found]}$ executes the command $\cde{C0}$ specified in
$\cde{watch(C0, B1)}$ if the constraint $\cde{B1}$ is entailed by the
store of the space the robot is in, i.e., an unwanted message is
found.

\begin{maude}
	crl [found] :
	  < L0 : agent | store : B0, set : SN >
	  < L0 : process | UID : I0, command : watch(C0, B1) >
	  < I : simulation | pqueue : T(Ra,((Ti, I0)), Le, Ri), nextID : N,
	    flag : false, pend : P, counter : N1, tTM : TMt, sTM : TMs, eTM : TMe, Atts >
	=>
	  < L0 : agent | store : B0, set : SN >
	  < L0 : process | UID : N, command : C0 >
	  < I : simulation | pqueue : T(Ra,((Ti, I0)), Le, Ri), nextID : (N + 1),
	    flag : true, pend : H0, counter : N2, tTM : TMt, sTM : TMs, eTM : TMe, Atts >
	if  entails(B0, B1)
	/\ (T0, N2) := getTimeCmd(C0, L0, TMt, TMs, TMe, N1)
	/\ H0 := insert(((T0, N)), P) .
\end{maude}

\noindent In this case study, the command $\cde{C0}$ represents
posting a warning message (i.e., posting a specific constraint to the
local store). However, the rule can still be used if other commands
were to be used. For intance, if the goal is to keep searching for
unwanted messages, the command $\cde{C0}$ could correspond to
$\cde{watch(C0, B1) || tell(`...')}$, where the command tell marks the
current store as containing an unwanted message.

\subsection{Using $\PVESTA$}






The initial state of the $\SSCC$ system is created by the
$\cde{initState}$ function, where the seed for sampling ($\cde{N}$ as
$\cde{Nat}$) is included. Besides the initial structure and the
process (including the robot process), a $\cde{flg}$ message is placed
in the initial configuration for indicating whether the simulation
will continue (as $\cde{Bool}$) and the maximum time for the
simulation (as $\cde{Float}$).


\begin{maude}
	---- init state
	op initState : Nat -> Configuration .
	op initState : -> Configuration .
	rl initState => initState(counter) .
	
	eq initState(N) 
	=  ... flg(false, 0.0) .  
\end{maude}

The property to be checked is the expected time it takes the robot
find an unwanted message. This is achieved with the help of the
$\cde{val(\_,\_)}$ function: it takes as arguments a unique natural
number to identify the property (as $\cde{Nat}$) and the configuration
of the system (as $\cde{Configuration}$). The function returns the
corresponding value of the property (as $\cde{Float}$). The $\QUATEX$
expression uses the $\cde{val(\_,\_)}$ function to calculate
$\mathbf{E}[Exp]$.

\begin{maude}
	op val : Nat Configuration -> Float .
	eq val(1, Conf) = getExcecutionTime(Conf) .
\end{maude}

The expected execution time of the sample is taken from the
$\cde{gtime}$ argument of the $\cde{simulation}$ object in the
configuration. Note that such a duration is parametric on the random
sampling of the stochastic expressions defined in the time functions
(e.g., $\alpha, \mu, \phi, \rho$) defined as part of the initial state
of the system.

\begin{maude}
	--- execution time
	op getExcecutionTime : Configuration -> Float .
	eq getExcecutionTime(< I : simulation | gtime : T, Atts > Conf)
	= float(T) .
\end{maude}

%

Finally, given the unique identifier of the property assigned by the 
$\cde{val(\_,\_)}$ function, the $\QUATEX$ expression to be checked is defined 
as \[\cde{execTime ( ) = { s.rval( 1 ) } ;}\] where $\cde{s.rval( 1 )}$ 
command connects $\PVESTA$ with the probabilistic system. Then, the expression 
is evaluated with the command \[\cde{eval E[ \# execTime ( ) ] ;}\] which 
output is the expected value of the property.

\subsection{Experimentation}

The setup of the experiments is the following: a hierarchical
structure of spaces is generated at random (up to some given depth),
with constraints and processes. The robot process is included in the
initial configuration of processes. Note that other processes can
concurrently modify the contents of any store, so that it makes sense
that the robot can revisit a space.

The experiment consisted in computing the expected value of the time
it takes the robot to find an unwanted post in different hierarchical
configurations. In particular, these configurations are generated from
depth 5 to depth 13 at random. The results are presented in
Table~\ref{tab:res-pvesta}; the experimentation was performed in a
cluster with a front-end with 16 cores and 32 GB of RAM, and four
workers with 64 cores and 64 GB of RAM each. The experiment with
height 13 did not finished because of memory issues.

\begin{table}[hbtp]
	\centering
  {\small
	\begin{tabular}{|cccccc|}
		\hline
		Hierarchy & Spaces &  Samples & RAM & Exec. time & Expected value \\
		(depth) & (count) &  (count) & (GB) & (sec) & (units) \\
		\hline\hline
		5 & 11 & 2400 & 6.60 & 175.19 & 38.88 \\
		6 & 23 & 1800 & 22.60 & 770.27 & 201.94 \\
		7 & 42 & 300 & 6.00 & 282.45 & 324.11 \\
		8 & 48 & 1800 & 12.90 & 605.51 & 111.74 \\
		9 & 65 & 300 & 12.20 & 895.12 & 676.60 \\
		10 & 81 & 600 & 10.60 & 1024.89 & 303.05 \\
		11 & 97 & 300 & 12.50 & 1541.57 & 716.75 \\
		12 & 115 & 300 & 57.50 & 11607.78 & 3312.83 \\
		\rowcolor{lightgray}
		13 & 144 & -- & 120.00* & 25200.00* & -- \\ \hline
	\end{tabular}
  }
	\caption{Expected values computed by the robot case study. From left
    to right, the columns represent the height of the search tree, the
    number of spaces in the tree, the amount of memory RAM used for
    the execution, the run-time of the execution, and expected time it
    takes the robot find an unwanted post.}
	\label{tab:res-pvesta}
\end{table}

It is important to note that $\PVESTA$ performs different number of
runs to compute the expected value of the robot finding the unwanted
message, depending on the height of the tree and the number of
nodes. On a separate note, this experimentation suggests that
$\PVESTA$ uses a significant amount of memory, even if the number of
spaces is small.





\section{Related Work}
\label{sec.related}

Distributed information is a central notion in systems with
hierarchical structure, where agents hold spaces with data and
processes.  S. Knight et al.~\cite{knight-sccp-2012} extend the theory
of $\CCP$ with spatial distribution of information in the
\emph{spatial constraint system} ($\SCS$). Computational hierarchical
spaces can be assigned to belong to agents, and each space may have
$\CCP$ processes and other (sub) spaces. Processes can post and query
information in their given space (i.e., locally) and may move from one
space to another.  M. Guzman et al.~\cite{guzman-sccpe-2017} develop
the theory of spatial constraint systems to provide a mechanism for
specifying the mobility of information or processes from one space to
another.  In~\cite{guzman-riscc-2018}, spatial constraint systems are
used as an abstract representation of modal logics.  This is useful to
characterize the notion of normality for self-maps in a constraint
system. As a counterpart of normal modal operator, it is shown that a
self-map is normal if and only if it preserves finite suprema. Also,
this abstraction is used to derive right inverse operators for modal
languages such as Kripke spatial constraint systems.  D. Gilbert and
C. Palamidessi~\cite{gilbert-ccpmob-2000} propose a different approach
to characterize process mobility using labeled transition systems.

Other efforts have focused on extending concurrent constraint
programming with temporal behavior. For example, V. Saraswat,
R. Jagadeesan, and V. Gupta~\cite{saraswat-tcc-1994} present a timed
asynchronous computation model and propose an implementation using
loop-free deterministic finite automata, a declarative framework for
reactive systems where time is represented as discrete time units.
They also present a non-deterministic version of $\CCP$
in~\cite{saraswat-deftcc-1995}, extending the model to express strong
time-outs and preemption.  Jagadeesan et
al.~\cite{jagadeesan-tccpctrl-2005} propose a policy algebra in the
timed concurrent constraint programming paradigm that uses a form of
default constraint programming and reactive computing to deal with
explicit denial, inheritance, overriding, and history-sensitive access
control.  M. Nielsen et al.~\cite{nielsen-temporal-2002} introduce a
model of temporal concurrent constraint programming, adding the
capability of modeling asynchronous and nondeterministic timed
behavior.  Additionally, they propose a proof system for
linear-temporal properties.  G. Sarria and
C. Rueda~\cite{sarria-rtcc-2008} present an extension of
$\textsf{ntcc}$ for specifying and modeling real-time behavior; their
operational semantics supports resources and limited time, and define
a denotational semantics.  As an application, the formal modeling of a
music improvisation example is implemented in this language.  F. de
Boer et al.~\cite{deboer-timedcc-2000} define a timed extension of
$\CCP$ with more expressive power than $\CCP$.

Extensions of $\CCP$ with probability have also been explored.
In~\cite{gupta-pcc-1997}, discrete random variables are introduced
with a given probability distribution. A new operator defines the
probability distribution of a random variable that is used to select
whether a process is executed or precluded. Random variables may be
constrained, and thus inconsistencies may arise between the chosen
values of random variables and constraints in the store.  Those
inconsistencies can cause some system runs to be precluded.
In~\cite{perez-pntcc-2008}, an operational semantics is proposed by
using probabilistic automata, with the final goal of extending
$\textsf{tcc}$~\cite{saraswat-tcc-1994} with probabilistic and
non-deterministic choices for processes.  A probabilistic choice
operator is defined to select processes guarded by constrains and
determine which one will be executed. When a process is chosen for
execution, the other processes are blocked.  Also, the notion of {\em
  probabilistic eventuality} is introduced to model the possible delay
of a process $P$ to be executed.  If $r$ represents the probability
for executing $P$ in the current time unit, the closer $r$ is to 1,
the greater the probability of executing $P$ will be. Analogously, $1
- r$ denotes the probability of delaying the execution of $P$. In such
a case, the given probability distribution modifies $r$ in a recursive
call used to reserve $P$ for the next time unit.  In
~\cite{klin-sosspc-2008} labeled continuous time Markov chains are
used to provide the semantics of stochastic processes.  If $x$ and $y$
are states, and $a$ is a label to move from $x$ to $y$, the
exponential distribution governs the duration of the transition from
$x$ to $y$ with label $a$. Conditional probability is used to govern
the transition $x \stackrel{a}{\to} y$: it is the probability that $x$
makes the transition to $y$ by an $a$-transition.

The inclusion of stochastic information for processes has also been
proposed. J. Aranda et al.~\cite{aranda-pntccsos-2008} associate a
random variable to each computation for determining its time duration:
given a set of competing actions, the fastest action is executed
(i.e., the one with the shortest duration). Based on $\CCP$,
D. Chiarugi et al.~\cite{chiarugi-stochmodel-2013} implement a
technique for the stochastic simulation of biochemical reactions with
non-Markovian behavior. V. Gupta et al.~\cite{gupta-stochastic-1999}
describe a stochastic concurrent constraint language for the
description and programming of concurrent probabilistic systems. In
this language, programs encode probability distributions over sets of
objects. Also, structural operational semantics and denotational
semantics are provided.

Finally, in the realm of rewriting logic, some executable semantics in
the Maude system have been proposed for $\CCP$-based models. M. Romero
and C. Rocha~\cite{romero-sccptr-2018} present a symbolic rewriting
logic semantics of the spatial modality of $\CCP$ with extrusion based
on the work in~\cite{knight-sccp-2012,guzman-sccpe-2017}. More
recently, M. Romero and C. Rocha~\cite{romero-sccpnfm-2018} have
proposed a symbolic rewriting logic semantics of the spatial modality
of $\CCP$ with extrusion and real-timing behavior. Somewhat related,
P. Degano et al.~\cite{degano-ccs-2002} provide a rewriting logic
semantics for Milner's CCS with interleaving behavior. Additionally, a
set of axioms is defined for a logical characterization of the
concurrency of CCS processes.

\section{Concluding Remarks}
\label{sec.concl}

This paper has presented a rewriting logic semantics for $\SSCC$, a
probabilistic, timed, and spatial concurrent constraint model. It is
fully executable in the Maude system.  The intended models of $\SSCC$
are spatially-distributed multi-agent reactive systems that may have
different computing capabilities, and be subject to real-time
requirements and probabilistic choice. In this setting, time
attributes are associated to process-store interaction, as well as to
process mobility in the space structure, by means of maps from agents
to probability distribution functions. Details about the underlying
constraint system have been given as materialized with the help of
rewriting modulo SMT and real-time behavior with the help of Real-Time
Maude. Furthermore, examples of quantitative analysis based on
statistical model checking have been given to illustrate key features
of $\SSCC$.

Future work includes the study of a structural operational semantics
for $\SSCC$ and proofs of correspondence between this new development
and the rewriting logic semantics contributed by this work. Also,
algebraic properties of $\SSCC$ need to be explored and, if possible,
establish relationships with previous extensions of $\CCP$ with time,
probabilities, and spaces. Finally, new case studies need to be
developed for $\SSCC$.

\section*{References}

\bibliographystyle{abbrv}
\bibliography{biblio}

\newpage

\appendix

\section{Scaffolding}
\label{sec.rew.scaffolding}

The rewriting logic semantics is specified in Maude as a collection of
modules and module operations presented in~\ref{a.appendix}.

\subsection{Leftist Heap}
\label{sec.heap}

The run-to-completion time of commands is simulated with the help of a
leftist heap that keeps track of all the active commands that are
waiting for the global timer to advance. The processes in the heap are
represented by pairs $(N, t)$ where $N$ is a unique identifier and
$t$ is its execution time. In any initial state of computation, all
processes are added to the heap and they are sorted with respect to
their execution time. A process is executed when its execution time is
the least time of all the processes that are pending to complete their
transitions.  When a process $(N, t)$ is at the root of the heap, it
is removed and the execution time of the remaining processes in the
heap is reduced $t$ units. When execution ends, the heap will be empty
or will contain only processes that can not make any further
transition.

A leftist heap~\cite{okasaki-datastructures-2003} is a heap-ordered
binary tree that satisfies the \emph{leftist property}: the rank of
any left child is at least as large as the rank of its right
sibling. It is implemented as a parametrized container in the
functional module
$\cde{LEFTIST-HEAP\{X~::~STRICT-TOTAL-ORDER\}}$. Admissible sets of
elements are strict totally ordered sets.  Heaps are constructed from
the constant $\cde{empty}$ and by means of the constructor operator
written as $\cde{T(\_,\_,\_,\_)}$. For a given heap $\cde{T(Ra,E,L,R)}$,
$\cde{E}$ is a node element, $\cde{Ra}$ is the rank of $\cde{E}$; and
$\cde{L}$ and $\cde{R}$ are the left and right siblings of $\cde{E}$,
respectively. Auxiliary operations include $\cde{isEmpty}$, $\cde{rank}$,
and $\cde{makeT}$, which are used to verify whether a heap is empty,
compute the rank of a given heap, and create a heap out of two heaps,
respectively.

\begin{maude}
  op isEmpty : Heap{X} -> Bool .
  op rank    : Heap{X} -> Nat .
  op makeT   : XElt Heap{X} Heap{X} -> NeHeap{X} .

  eq isEmpty(empty)       = true .
  eq isEmpty(T(Ra,E,L,R)) = false .
  eq rank(empty)       = 0 .
  eq rank(T(Ra,E,L,R)) = Ra .
  eq makeT(E,L,R)
   = if rank(L) >= rank(R)
     then T(rank(R) + 1,E,L,R)
     else T(rank(L) + 1,E,R,L)
     fi .
  \end{maude}
  Heap operations are defined as follows:
  \begin{maude}
  op findMin   : NeHeap{X} -> X\$Elt .
  op deleteMin : NeHeap{X} -> Heap{X} .
  op insert    : X\$Elt Heap{X} -> NeHeap{X} .
  op merge     : Heap{X} Heap{X} -> Heap{X} .
  
  eq findMin(T(Ra,E,L,R)) = E .
  eq deleteMin(T(Ra,E,L,R)) = merge(L,R) .
  eq insert(E,L) = merge(T(1,E,empty,empty),L) .
  eq merge(empty, L) = L .
  eq merge(L, empty) = L .
  eq merge(T(Ra,E,L,R),T(Ra',E',L',R')) 
   = if (E < E')
     then makeT(E,L,merge(R,T(Ra',E',L',R')))
     else makeT(E',L',merge(T(Ra,E,L,R),R'))
     fi .
  \end{maude}
  The length of the left spine of any node is always at
  least as large as the length of its right spine because of the
  condition $\cde{(E < E')}$ in equation for $\cde{merge}$. Operations
  $\cde{insert}$ and $\cde{deleteMin}$ are based on the merge operation;
  $\cde{findMin}$ returns the element at the top of a non-empty heap.

\subsection{SMT Solver}

SMT solving technology available from the current version of Maude is
used to realize the underlying constraint system of $\SSCC$. The
sort $\cde{Boolean}$ (available in the current version of Maude from
the $\cde{INTEGER}$ module) defines the data type used to represent
$\PTSCCP$'s constraints.  Terms of sort $\cde{Boolean}$ are
quantifier-free formulas built from variables ranging over the
Booleans and integers, and the usual function symbols. The current
version of Maude is integrated with the CVC4~\cite{barrett-cvc4-2011}
and Yices2~\cite{dutertre-yices2-2014} SMT solvers, which can be
queried via the meta-level. In this semantics, queries to the SMT
solvers are encapsulated by functions $\cde{check-sat}$ and
$\cde{check-unsat}$:

\begin{maude}
  op check-sat : Boolean -> Bool .
  op check-unsat : Boolean -> Bool .
  eq check-sat(B)
   = metaCheck(['INTEGER], upTerm(B)) .
  eq check-unsat(B)
   = not(check-sat(B)) .
\end{maude}

\noindent The function invocation $\cde{check-sat}(B)$ returns true
only if $B$ is satisfiable. Otherwise, it returns false if it
unsatisfiable or undefined if the SMT solver cannot decide. Note that
function invocation $\cde{check-unsat}(B)$ returns true only if $B$ is
unsatisfiable. Therefore, the rewriting logic semantics of $\PTSCCP$
instantiates the constraint system $(\textit{Con},\sqsubseteq)$ by
having quantifier-free formulas as the constraints $\textit{Con}$ and
semantic validity (w.r.t. the initial model of the theory queried in
the SMT solver) as the entailment relation $\sqsubseteq$. More
precisely, if $\Gamma$ is a finite set of terms of sort
$\cde{Boolean}$ and $\phi$ is term of sort $\cde{Boolean}$, the
following equivalence holds:
\[\Gamma \sqsubseteq \phi \quad \iff \quad
\cde{check-unsat}\left(\left(\bigwedge_{\gamma\in\Gamma} \gamma\right) \land
\neg\phi\right).\]

In order to make a direct relation between the entailment relation
$\sqsubseteq$ and the Maude syntax, the operator $\cde{entails}$ is
defined as follows:

\begin{maude}
  op entails : Boolean Boolean -> Bool .

  eq entails(C1:Boolean, C2:Boolean)
   = check-unsat(C1:Boolean and not(C2:Boolean)) .
\end{maude}
This operator returns true if $\cde{C2}$ can be derived from $\cde{C1}$.

\subsection{The tick rule}
\label{sec.rew.tick}

\begin{figure}
	\centering
	\resizebox{7cm}{!}{%
		\begin{tikzpicture}[node distance=3cm,
		every node/.style={fill=white, font=\rmfamily\footnotesize}, align=center,
		auto]
		\node (pqueue)  [state] {\textit{pqueue}};
		\node (pending)	  [state, left of=root]	{\textit{pending}}
		edge[pil,->, bend left=25] node[sloped,above,yshift=3pt] {\rlname{tick}}
		(pqueue)
		edge[pil,<-, bend left=-25] node[sloped,below,yshift=-3pt] {\rlname{delay}}
		(pqueue);
		\node (exec)  	  [state, right of=root]
		{\textit{exec}};
		\draw[->]   (pqueue) -- (exec) node[midway,sloped,above] {\rlname{ask}};
		\end{tikzpicture}}
	\caption{Possible transitions for ask commands.}
	\label{fig.tick.pending}
\end{figure}
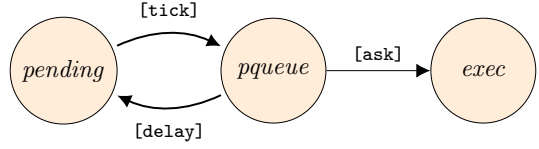

A real-time rewrite theory is a rewrite theory where some rules, called tick 
rules, model time elapse in the system~\cite{olveczky-realtime-2002}.
Here the tick rule is defined as follows:

\begin{maude}
 crl [tick] :
     < I : simulation | pqueue : P, gtime : T, flag : true, pend : P0, Atts >
  => < I : simulation | pqueue : merge(delta(deleteMin(P),T0),P0), gtime :  (T plus T0), 
                        flag : false, pend : empty, Atts >
  if T0 := p1(findMin(P)) .
\end{maude}
where the auxiliary operation $\cde{delta}$ reduces $\cde{T0}$
units the execution time of every command in the heap $\cde{P}$:
\begin{maude}
  op delta : Heap{2Tuple} Time -> Heap{2Tuple} .
  eq delta(empty,T') = empty .
  eq delta(T(N,((T1, I)),P,P0),T') = T(N,((T1 monus T', I)),delta(P,T'),delta(P0,T')) .
\end{maude}
When the $\rlname{tick}$ rule is fired, the global time $\cde{T}$ is
incremented in $\cde{T0}$ units, where $\cde{T0}$ is the minimum time
present in the priority queue $\cde{P}$, which is modified by removing
the process with the minimum execution time. It also adds the pending
commands in heap $\cde{P0}$ to the priority queue $\cde{P}$. The
pending commands are querying commands that, although they have been
activated already for execution, have not yet been able to execute because
their guard has not been entailed by the current state of the
corresponding local stores. The tick rule puts all these pending
process back in the main queue $\cde{P}$, so that their guards can be
checked again and be executed or put back in the pending
queue. Figure~\ref{fig.tick.pending} depicts the possible transitions
that an ask command can take between being in the priority queue, in
the pending queue, and finally executing.

\section{Rewriting Logic Semantics of $\SSCC$}
\label{a.appendix}

This appendix includes the specification in Maude of Real-Time $\SSCC$.

\subsection{agent-id.maude}

\begin{maude}
--- agent identifier
fmod AGENT-ID is
  pr EXT-BOOL .
  pr NAT .
  sort Aid .
  
  op root : -> Aid .
  op _._ : Nat Aid -> Aid .
  
  vars L L0 L1 L2 : Aid .
  vars N N0 N1 N2 : Nat .
  
  --- auxiliary operations
  op is-prefix? : Aid Aid -> Bool .
  eq is-prefix?(root, L) = true .
  eq is-prefix?(N . L, root) = false .
  eq is-prefix?(N0 . L0, N1 . L1) = (N0 . L0  == N1 . L1) or-else is-prefix?(N0 . L0, L1) .
  
  op is-son? : Aid Aid -> Bool .
  eq is-son?(root, L) = false .
  eq is-son?(N . L, L0) = (L == L0) .
  
  op sizeAid : Aid -> Nat .
  eq sizeAid(root) = 1 .
  eq sizeAid(N . L) = 1 + sizeAid(L) .
endfm
\end{maude}

\subsection{smt-util.maude}

\begin{maude}
fmod SMT-UTIL is
  inc INTEGER .
  pr CONVERSION .
  pr META-LEVEL .

  op check-sat : Boolean -> Bool .
  op check-unsat : Boolean -> Bool .
  op entails : Boolean Boolean -> Bool .
  eq check-sat(B:Boolean) = metaCheck(['INTEGER], upTerm(B:Boolean)) .
  eq check-unsat(B:Boolean) = not(check-sat(B:Boolean)) .
  eq entails(C1:Boolean, C2:Boolean) 
   = check-unsat(C1:Boolean and not(C2:Boolean)) .

  --- some Boolean identities
  eq B:Boolean and true = B:Boolean .
  eq B:Boolean and false = false .
  eq B:Boolean or true = true .
  eq B:Boolean or false = B:Boolean .
  eq true and B:Boolean = B:Boolean .
  eq false and B:Boolean = false .
  eq true or B:Boolean = true .
  eq false or B:Boolean = B:Boolean .
  eq not((true).Boolean) = (false).Boolean .
  eq not((false).Boolean) = (true).Boolean .
endfm
\end{maude}

\subsection{2-tuple.maude}

\begin{maude}
fmod 2-TUPLE{X :: STRICT-TOTAL-ORDER, Y :: STRICT-TOTAL-ORDER} is
  sort Tuple{X, Y} .

  op ((_,_)) : X\$Elt Y\$Elt -> Tuple{X, Y} [ctor] .
  op p1_ : Tuple{X, Y} -> X\$Elt .
  op p2_ : Tuple{X, Y} -> Y\$Elt .
  eq p1(A:X\$Elt, B:Y\$Elt) = A:X\$Elt .
  eq p2(A:X\$Elt, B:Y\$Elt) = B:Y\$Elt .

  op _<_ : Tuple{X, Y} Tuple{X, Y} -> Bool .
  eq A:Tuple{X, Y} < B:Tuple{X, Y} = p1(A:Tuple{X, Y}) < p1(B:Tuple{X, Y}) .
endfm

fmod PAIR{X :: TRIV, Y :: TRIV} is
  sort Pair{X, Y} .
  
  op P : X\$Elt Y\$Elt -> Pair{X, Y} [ctor] .
  op p1_ : Pair{X, Y} -> X\$Elt .
  op p2_ : Pair{X, Y} -> Y\$Elt .
  eq p1(P(A:X\$Elt, B:Y\$Elt)) = A:X\$Elt .
  eq p2(P(A:X\$Elt, B:Y\$Elt)) = B:Y\$Elt .
endfm
\end{maude}

\subsection{4-tuple.maude}

\begin{maude}
fmod 4-TUPLE{W :: TRIV, X :: TRIV, Y :: TRIV, Z :: TRIV} is
  sort Tuple{W, X, Y, Z} .

  op ((_,_,_,_)) : W\$Elt X\$Elt Y\$Elt Z\$Elt -> Tuple{W, X, Y, Z} [ctor] .
  op p1_ : Tuple{W, X, Y, Z} -> W\$Elt .
  op p2_ : Tuple{W, X, Y, Z} -> X\$Elt .
  op p3_ : Tuple{W, X, Y, Z} -> Y\$Elt .
  op p4_ : Tuple{W, X, Y, Z} -> Z\$Elt .
  eq p1(A:W\$Elt, B:X\$Elt, C:Y\$Elt, D:Z\$Elt) = A:W\$Elt .
  eq p2(A:W\$Elt, B:X\$Elt, C:Y\$Elt, D:Z\$Elt) = B:X\$Elt .
  eq p3(A:W\$Elt, B:X\$Elt, C:Y\$Elt, D:Z\$Elt) = C:Y\$Elt .
  eq p4(A:W\$Elt, B:X\$Elt, C:Y\$Elt, D:Z\$Elt) = D:Z\$Elt .
endfm
\end{maude}

\subsection{leftistHeap.maude}

\begin{maude}
fmod LEFTIST-HEAP{X :: STRICT-TOTAL-ORDER} is
  protecting NAT .
  sort Heap{X} NeHeap{X} .
  subsort NeHeap{X} < Heap{X} .

  op empty : -> Heap{X} .
  op T(_,_,_,_) : Nat X\$Elt Heap{X} Heap{X} -> NeHeap{X} .

  vars L L' R R' : Heap{X} .
  vars E E'       : X\$Elt .
  vars Ra Ra'     : Nat .

  op isEmpty : Heap{X} -> Bool .
  eq isEmpty(empty) = true .
  eq isEmpty(T(Ra,E,L,R)) = false .

  op rank : Heap{X} -> Nat .
  eq rank(empty) = 0 .
  eq rank(T(Ra,E,L,R)) = Ra .

  op makeT : X\$Elt Heap{X} Heap{X} -> NeHeap{X} .
  eq makeT(E,L,R)
   = if rank(L) >= rank(R)
     then T(rank(R) + 1,E,L,R)
     else T(rank(L) + 1,E,R,L)
     fi .

  op merge : Heap{X} Heap{X} -> Heap{X} .
  eq merge(empty, L) = L .
  eq merge(L, empty) = L .
  eq merge(T(Ra,E,L,R),T(Ra',E',L',R'))
   = if (E < E')
     then makeT(E,L,merge(R,T(Ra',E',L',R')))
     else makeT(E',L',merge(T(Ra,E,L,R),R'))
     fi .

  op insert : X\$Elt Heap{X} -> NeHeap{X} .
  eq insert(E,L) = merge(T(1,E,empty,empty),L) .

  op findMin : NeHeap{X} -> X\$Elt .
  eq findMin(T(Ra,E,L,R)) = E .

  op deleteMin : NeHeap{X} -> Heap{X} .
  eq deleteMin(T(Ra,E,L,R)) = merge(L,R) .
endfm
\end{maude}

\subsection{stochastic-expression.maude}

\begin{maude}
view Time from STRICT-TOTAL-ORDER to POSRAT-TIME-DOMAIN is
  sort Elt to Time .
endv

view 2Tuple from STRICT-TOTAL-ORDER to 2-TUPLE{Time, Nat<} is
  sort Elt to Tuple{Time, Nat<} .
endv

fmod STOCHASTIC-EXPRESSION is
  pr MGDISTRIBUTIONS .
  pr POSRAT-TIME-DOMAIN .  
  pr 2-TUPLE{Time, Nat<} .
  pr 2-TUPLE{Float<, Nat<} .
  
  sort StochasticExpression .
  subsort Time < StochasticExpression .
  
  vars F L M SG SH SC ND DD P LB UB : Float .
  vars S LI MI SGI SHI SCI NDI DDI PI LBI UBI : Nat .
  vars LPR MPR SGPR SHPR SCPR NDPR DDPR PPR LBPR UBPR : PosRat .
  var  T : Time .
  
  ---- NORMAL/GAUSS DISTRIBUTION  
  op Norm : -> StochasticExpression .
  op Norm : Float Float -> StochasticExpression .
  eq Norm = Norm(0.0, 1.0) .
  
  ---- EXPONENTIAL DISTRIBUTION
  op Exp : Float -> StochasticExpression .
  
  ---- UNIFORM DISTRIBUTION
  op Unif : Float Float -> StochasticExpression .
  
  ---- GAMMA DISTRIBUTION
  op Gam : Float Float -> StochasticExpression .
  
  ---- WEIBULL DISTRIBUTION
  op Weib : Float Float -> StochasticExpression .
  
  ---- CHI-SQUARE DISTRIBUTION
  op Chi : Float -> StochasticExpression .
  
  ---- LOG-NORMAL DISTRIBUTION
  op Log : Float Float -> StochasticExpression .
  
  op eval : StochasticExpression Int -> Tuple{Time, Nat<} .
  eq eval(Norm(M, SG), S)
   = (if normDistr(M, SG, S) >= 0.0 
      then rat(normDistr(M, SG, S)) 
      else 0 
      fi, s S) .
  eq eval(Exp(L), S)
   = (if expDistr(L, S) >= 0.0 
      then rat(expDistr(L, S)) 
      else 0 
      fi, s S) .
  eq eval(Unif(LB, UB), S)
   = (if unifDistr(LB, UB, S) > 0.0 
      then rat(unifDistr(LB, UB, S)) 
      else 0 
      fi, s S) .
  eq eval(Gam(SH, SC), S)
   = (if gammaDistr(SH, SC, S) > 0.0 
      then rat(gammaDistr(SH, SC, S)) 
      else 0 
      fi, s S) .
  eq eval(Weib(SC, SH), S)
   = (if weibDistr(SC, SH, S) > 0.0 
      then rat(weibDistr(SC, SH, S)) 
      else 0 
      fi, s S) .
  eq eval(Chi(SH), S)
   = (if chiSDistr(SH, S) > 0.0 
      then rat(chiSDistr(SH, S)) 
      else 0 
      fi, s S) .
  eq eval(Log(M, SG), S)
   = (if logNormDistr(M, SG, S) > 0.0 
      then rat(logNormDistr(M, SG, S)) 
      else 0 
      fi, s S) .
  eq eval(T, S) = (T, S) .
  
  op evalF : StochasticExpression Int -> Tuple{Float<, Nat<} .
  eq evalF(Unif(LB, UB), S) = (unifDistr(LB, UB, S), s S) .
endfm
\end{maude}

\subsection{sccp.maude}

  \begin{maude}
--- commands syntax
fmod SCCP-SYNTAX is
  pr INTEGER .
  pr AGENT-ID .
  
  sort SCCPCmd .
  
  op 0       :                 -> SCCPCmd .
  op tell    : Boolean         -> SCCPCmd .
  op ask_->_ : Boolean SCCPCmd -> SCCPCmd .
  op _||_    : SCCPCmd SCCPCmd -> SCCPCmd [assoc comm gather (e E) ] .
  op _in_    : SCCPCmd Nat     -> SCCPCmd .
  op _out_   : SCCPCmd Nat     -> SCCPCmd .
  
  op V       : Nat             -> SCCPCmd .
  op mu      : Nat     SCCPCmd -> SCCPCmd .
endfm

view SCCPCmd from TRIV to SCCP-SYNTAX is
  sort Elt to SCCPCmd .
endv

view Aid from TRIV to AGENT-ID is
  sort Elt to Aid .
endv

fmod SCCP-SYNTAX-EXT is
  pr SCCP-SYNTAX .
  pr LIST{SCCPCmd} .
  pr LIST{Float} .
  pr ALIST{Aid} .

  op exc : List{SCCPCmd} List{Float} -> SCCPCmd .
  op ind : List{SCCPCmd} List{Float} -> SCCPCmd .

  op watch : SCCPCmd Boolean -> SCCPCmd .
endfm

view Aid from TRIV to AGENT-ID is
  sort Elt to Aid .
endv

view StExp from TRIV to STOCHASTIC-EXPRESSION is
  sort Elt to StochasticExpression .
endv

--- state syntax
mod SCCP-STATE is
  pr SCCP-SYNTAX-EXT .
  inc CONFIGURATION .
  pr STOCHASTIC-EXPRESSION .
  pr LEFTIST-HEAP{2Tuple} .
  pr MAP{Aid, StExp} .
  pr SET{Nat} * (op _,_ to _;_) .
  
  sort Sys .
  subsorts Nat Aid < Oid .
  
  ops agent process simulation :    -> Cid .
  
  op store :_    : Boolean          -> Attribute [ctor] .
  op set :_      : Set{Nat}         -> Attribute [ctor] .
  op UID :_      : Nat              -> Attribute [ctor] .
  op command :_  : SCCPCmd          -> Attribute [ctor] .
  op gtime :_    : Time             -> Attribute [ctor] .
  op pqueue :_   : Heap{2Tuple}     -> Attribute [ctor] .
  op pend :_     : Heap{2Tuple}     -> Attribute [ctor] .
  op nextID :_   : Nat              -> Attribute [ctor] .
  op counter :_  : Nat              -> Attribute [ctor] .
  op flag :_     : Bool             -> Attribute [ctor] .
  op timeMap :_  : Map{Aid, StExp}  -> Attribute [ctor] .
  op tTM :_      : Map{Aid, StExp}  -> Attribute [ctor] .
  op aTM :_      : Map{Aid, StExp}  -> Attribute [ctor] .
  op sTM :_      : Map{Aid, StExp}  -> Attribute [ctor] .
  op eTM :_      : Map{Aid, StExp}  -> Attribute [ctor] .
  op factor :_   : PosRat           -> Attribute [ctor] .
endm

view List from TRIV to LIST{SCCPCmd} is
  sort Elt to List{SCCPCmd} .
endv

view Heap from TRIV to LEFTIST-HEAP{2Tuple} is
  sort Elt to Heap{2Tuple} .
endv

view Map from TRIV to MAP{Aid, StExp} is
  sort Elt to Map{Aid, StExp} .
endv

view FList from TRIV to LIST{Float} is
  sort Elt to List{Float} .
endv

--- transitions
mod SCCP is
  inc SCCP-STATE .
  pr SMT-UTIL .
  pr 4-TUPLE{List, Nat, Heap, Nat} .
  pr PAIR{FList, Nat} .
  
  vars N N0 N1 N2 N3 N4 N5 S N' N'' N2' I I0 I1 Ra  : Nat .
  vars M M'          : NzNat .
  vars SN, SN'       : Set{Nat} .
  vars In1 In2 In3   : Integer .
  vars Q Q' Q1 Q1'   : Float .
  vars alpha beta    : PosRat .
  vars Bl Bl'        : Bool .
  vars T T' T0 T1 Ti : Time .
  vars L L0 L1       : Aid .
  vars B B0 B1       : Boolean .
  vars C C0 C1 C2    : SCCPCmd .
  vars H H' H0 H0' H1 H1'  : NeHeap{2Tuple} .
  vars P P' P0 P0' Le Ri   : Heap{2Tuple} .
  var  Atts          : AttributeSet .
  vars X             : Configuration .
  vars O             : Object .
  vars LC LC'        : List{SCCPCmd} .
  vars NeLC NeLC'    : NeList{SCCPCmd} .
  vars LF LF' F1     : List{Float} .
  vars TM TM' TMa TMt TMe TMs TMt' TMa' TMs' TMe'  : Map{Aid, StExp} .
    
  --- time delta 
  op delta : Heap{2Tuple} Time -> Heap{2Tuple} .
  
  eq delta(empty,T') = empty .
  eq delta(T(N,((T1, I)),P,P0),T') = T(N,((T1 monus T', I)),delta(P,T'),delta(P0,T')) .
  
  --- auxiliary operations  
  op get-ancestor : Map{Aid, StExp} Aid   -> StochasticExpression .
  eq get-ancestor(TM, root)
   = if $hasMapping(TM, root) 
     then TM[root] 
     else Norm(1.0, 0.2) 
     fi .
  eq get-ancestor(TM, N . L)
   = if $hasMapping(TM, N . L) 
     then TM[N . L] 
     else get-ancestor(TM, L) 
     fi .
  
  op size : Boolean -> Nat .
  eq size(not B0) = 1 + size(B0) .
  eq size(B0 and B1) = 1 + size(B0) + size(B1) .
  eq size(B0 xor B1) = 1 + size(B0) + size(B1) .
  eq size(B0 or B1) = 1 + size(B0) + size(B1) .
  eq size(B0 implies B1) = 1 + size(B0) + size(B1) .
  eq size(B0 === B1) = 1 + size(B0) + size(B1) .
  eq size(B0 =/== B1) = 1 + size(B0) + size(B1) .
  eq size(In1 < In2) = 1 .
  eq size(In1 <= In2) = 1 .
  eq size(In1 > In2) = 1 .
  eq size(In1 >= In2) = 1 .
  eq size(In1 === In2) = 1 .
  eq size(In1 =/== In2) = 1 .
  eq size(B0) = 1 [owise] .
  
  op replace : Nat SCCPCmd SCCPCmd -> SCCPCmd .
  eq replace(N, 0, C) = 0 .
  eq replace(N, tell (B), C) = tell (B) .
  eq replace(N, ask B -> C0, C) = ask B -> replace(N, C0, C) .
  eq replace(N, C0 || C1, C) = replace(N, C0, C) || replace(N, C1, C) .
  eq replace(N, exc(LC, LF), C) = exc(repalceInList(N, LC, C), LF) .
  eq replace(N, ind(LC, LF), C) = ind( repalceInList(N, LC, C), LF) .
  eq replace(N, C0 in N0, C) = replace(N, C0, C) in N0 .
  eq replace(N, C0 out N0, C) = replace(N, C0, C) out N0 .
  eq replace(N, mu(N0, C0), C) = mu(N0, C0) .
  eq replace(N, V(N0), C)
   = if (N0 == N) 
     then C 
     else V(N0) 
     fi .
  
  op repalceInList : Nat List{SCCPCmd} SCCPCmd -> List{SCCPCmd} .
  eq repalceInList(N, C0 LC, C) = replace(N, C0, C) LC .
  eq repalceInList(N, nil, C) = nil .
  
  --- get time from maps
  
  op fTime : Map{Aid, StExp} Aid Nat -> Tuple{Time, Nat<} .
  eq fTime(TM, L, N)
   = if hasMapping(TM, L) 
     then eval(TM[L], N) 
     else eval(Norm(1.0, 0.2), N) 
     fi .
  
  op getTimeCmd : SCCPCmd Aid Map{Aid, StExp} Map{Aid, StExp} Map{Aid, StExp} Nat -> Tuple{Time, Nat<} .
  eq getTimeCmd(tell(B1), L, TMt, TMs, TMe, N) = fTime(TMt, L, N) .
  eq getTimeCmd(C1 in I1, L, TMt, TMs, TMe, N) = fTime(TMs, L, N) .
  eq getTimeCmd(C1 out I1, L, TMt, TMs, TMe, N) = fTime(TMe, L, N) .
  eq getTimeCmd(C1, L, TMt, TMs, TMe, N) = (0, N) [owise] .
  
  op getProb : Nat -> Tuple{Float<, Nat<} .
  eq getProb(N2) = evalF(Unif(0.0, 1.0), N2) .
  
  --- exclusive and independent parallel functions
  op exclusive : List{SCCPCmd} List{Float} Nat Heap{2Tuple} Nat Map{Aid, StExp} Map{Aid, StExp} Map{Aid, StExp} Aid 
                 -> Tuple{List, Nat, Heap, Nat} .
 ceq exclusive(C, Q, N, P, N1, TMt, TMs, TMe, L) = (C, N + 1, H0, N')
  if (T, N') := getTimeCmd(C, L, TMt, TMs, TMe, N1) /\ H0 := insert(((T, N)), P) .
 ceq exclusive(C NeLC, Q Q1 LF, N, P, N1, TMt, TMs, TMe, L)
   = if Q' <= Q
     then (C, N + 1, H0, N'')
     else exclusive(NeLC, (Q + Q1) LF, N, P, N'', TMt, TMs, TMe, L)
     fi
  if (Q', N') := getProb(N1) /\ (T, N'') := getTimeCmd(C, L, TMt, TMs, TMe, N') /\ H0 := insert(((T, N)), P) .
  
  op independent : List{SCCPCmd} List{Float} List{SCCPCmd} Nat Heap{2Tuple} Nat Map{Aid, StExp} Map{Aid, StExp} 
                   Map{Aid, StExp} Aid -> Tuple{List, Nat, Heap, Nat} .
  eq independent(nil, nil, LC', N, P, N1, TMt, TMs, TMe, L) = (LC', N, P, N1) .
 ceq independent(C LC, Q LF, LC', N, P, N1, TMt, TMs, TMe, L)
   = if Q' <= Q
     then independent(LC, LF, LC' C, N + 1, H0, N'', TMt, TMs, TMe, L)
     else independent(LC, LF, LC', N, P, N'', TMt, TMs, TMe, L)
     fi
  if (Q', N') := getProb(N1) /\ (T, N'') := getTimeCmd(C, L, TMt, TMs, TMe, N') /\ H0 := insert(((T, N)), P) .
  
  op genCommands : List{SCCPCmd} Nat Aid -> Configuration .
  eq genCommands(nil, N, L) = none .
  eq genCommands(C LC, N, L) = < L : process | UID : N, command : C > genCommands(LC, N + 1, L) .
  
  op command-list : Aid Set{Nat} SCCPCmd -> List{SCCPCmd} .
  eq command-list(N . L, SN, C0) = (C0 out N) $command-list(SN, C0) .
  eq command-list(root, SN, C0) = $command-list(SN, C0) .
  
  op $command-list : Set{Nat} SCCPCmd -> List{SCCPCmd} .
  eq $command-list((N1 ; SN), C0) = (C0 in N1) $command-list(SN, C0) .
  eq $command-list(empty, C0) = nil .
  
  op prob-list : Nat Nat -> Pair{FList, Nat} .
  eq prob-list(N, N1) = $prob-list(N, N1, nil, 0.0) .
  
  op $prob-list : Nat Nat List{Float} Float -> Pair{FList, Nat} .
 ceq $prob-list(s N, N1, F1, Q1) = $prob-list(N, N', Q F1, Q1 + Q)
  if (Q, N') := getProb(N1) .
  eq $prob-list(0, N1, F1, Q1) = P($normalize(F1, Q1), N1) .
  
  op $normalize : List{Float} Float -> List{Float} .
  eq $normalize(Q F1, Q1) = (Q / Q1) $normalize(F1, Q1) .
  eq $normalize(nil, Q1) = nil .
  
  --- non-observable concurrent transitions
  
  eq < L0 : process | command : 0, Atts > = none .
  eq < L0 : agent | store : B0, set : SN > < L0 : agent | store : B1, set : SN' >
   = < L0 : agent | store : (B0 and B1), set : (SN ; SN') > .
  
  --- observable concurrent transitions
  
  rl [tell] :
     < L0 : agent | store : B0, set : SN > < L0 : process | UID : I0, command : tell (B1) >
     < I : simulation | pqueue : T(Ra,((Ti, I0)), Le, Ri), flag : false, Atts >
  => < L0 : agent | store : (B0 and B1), set : SN >
     < I : simulation | pqueue : T(Ra,((Ti, I0)), Le, Ri), flag : true, Atts >  .
  
  rl [tell-set] :
     < L0 : agent | store : B0, set : SN > < L0 : process | UID : I0, command : tell (N) >
     < I : simulation | pqueue : T(Ra,((Ti, I0)), Le, Ri), flag : false, Atts >
  => < L0 : agent | store : B0, set : (SN ; N) >
     < I : simulation | pqueue : T(Ra,((Ti, I0)), Le, Ri), flag : true, Atts >  .
  
 crl [parallel] :
     < L0 : process | UID : I0, command : (C0 || C1) >
     < I : simulation | pqueue : T(Ra,((Ti, I0)), Le, Ri), nextID : N, flag : false, pend : P, counter : N1, 
                        tTM : TMt, sTM : TMs, eTM : TMe, Atts >
  => < L0 : process | UID : N, command : C0 >
     < L0 : process | UID : (N + 1), command : C1 >
     < I : simulation | pqueue : T(Ra,((Ti, I0)), Le, Ri), nextID : (N + 2), flag : true, pend : H0, 
                        counter : N3, tTM : TMt, sTM : TMs, eTM : TMe, Atts >
  if (T0, N2) := getTimeCmd(C0, L0, TMt, TMs, TMe, N1) /\ (T1, N3) := getTimeCmd(C1, L0, TMt, TMs, TMe, N2) 
     /\ H0 := insert(((T0, N)), insert(((T1, N + 1)), P)) .
  
 crl [exclusive] :
     < L0 : process | UID : I0, command : exc(LC, LF) >
     < I : simulation | pqueue : T(Ra,((Ti, I0)), Le, Ri), nextID : N, flag : false, pend : P, counter : N1, 
                        tTM : TMt, sTM : TMs, eTM : TMe, Atts >
  => genCommands(LC', N, L0) 
     < I : simulation | pqueue : T(Ra,((Ti, I0)), Le, Ri), nextID : N', flag : true, pend : P0, counter : N'', 
                        tTM : TMt, sTM : TMs, eTM : TMe, Atts >
  if (LC', N', P0, N'') := exclusive(LC, LF, N, P, N1, TMt, TMs, TMe, L0) .
  
 crl [independent] :
     < L0 : process | UID : I0, command : ind(LC, LF) >
     < I : simulation | pqueue : T(Ra,((Ti, I0)), Le, Ri), nextID : N, flag : false, pend : P, counter : N1, 
                        tTM : TMt, sTM : TMs, eTM : TMe, Atts >
  => genCommands(LC', N, L0) 
     < I : simulation | pqueue : T(Ra,((Ti, I0)), Le, Ri), nextID : N', flag : true, pend : P0, counter : N'', 
                        tTM : TMt, sTM : TMs, eTM : TMe, Atts >
  if (LC', N', P0, N'') := independent(LC, LF, nil, N, P, N1, TMt, TMs, TMe, L0) .
  
 crl [space] :
     < L0 : agent | store : B0, set : SN >
     < L0 : process | UID : I0, command : (C0 in N0) >
     < I : simulation | pqueue : T(Ra,((Ti, I0)), Le, Ri), nextID : N, flag : false, pend : P, counter : N1, 
                        tTM : TMt, aTM : TMa, sTM : TMs, eTM : TMe, Atts >
  => < L0 : agent | store : B0, set : (SN ; N0) >
     < N0 . L0 : agent | store : true, set : empty >
     < N0 . L0 : process | UID : N, command : C0 >
     < L0 : process | UID : (N + 1), command : tell(N0) >
     < I : simulation | pqueue : T(Ra,((Ti, I0)), Le, Ri), flag : true, pend : H1, nextID : (N + 2), counter : N3, 
                        tTM : TMt', aTM : TMa', sTM : TMs', eTM : TMe', Atts >
  if TMt' := insert(N0 . L0, get-ancestor(TMt, N0 . L0), TMt) 
     /\ TMa' := insert(N0 . L0, get-ancestor(TMa, N0 . L0), TMa) 
     /\ TMs' := insert(N0 . L0, get-ancestor(TMs, N0 . L0), TMs) 
     /\ TMe' := insert(N0 . L0, get-ancestor(TMe, N0 . L0), TMe) 
     /\ (T0, N2) := getTimeCmd(C0, N0 . L0, TMt', TMs', TMe', N1) /\ H0 := insert(((T0, N)), P) 
     /\ (T1, N3) := getTimeCmd(tell(N0), L0, TMt', TMs', TMe', N2) /\ H1 := insert(((T1, (N + 1))), H0) .
  
 crl [extrusion]:
     < N0 . L0 : process | UID : I0, command : (C0 out N0) >
     < I : simulation | pqueue : T(Ra,((Ti, I0)), Le, Ri), nextID : N, flag : false, pend : P, counter : N1, 
                        tTM : TMt, sTM : TMs, eTM : TMe, Atts >
  => < L0 : process | UID : N, command : C0 >
     < I : simulation | pqueue : T(Ra,((Ti, I0)), Le, Ri), flag : true, pend : H0, nextID : (N + 1), counter : N2, 
                        tTM : TMt, sTM : TMs, eTM : TMe, Atts >
  if (T0, N2) := getTimeCmd(C0, L0, TMt, TMs, TMe, N1) /\ H0 := insert(((T0, N)), P) .
  
 crl [ask] :
     < L0 : agent | store : B0, set : SN >
     < L0 : process | UID : I0, command : (ask B1 -> C1) >
     < I : simulation | pqueue : T(Ra,((Ti, I0)), Le, Ri), flag : false, pend : P, nextID : N, counter : N1, 
                        tTM : TMt, aTM : TMa, sTM : TMs, eTM : TMe, factor : alpha, Atts >
  => < L0 : agent | store : B0, set : SN >
     < L0 : process | UID : N, command : C1 >
     < I : simulation | pqueue : T(Ra,((Ti, I0)), Le, Ri), flag : true, pend : H0, nextID : (N + 1), counter : N3, 
                        tTM : TMt, aTM : TMa, sTM : TMs, eTM : TMe, factor : alpha, Atts >
  if entails(B0,B1) /\ (T0, N2) := getTimeCmd(C1, L0, TMt, TMs, TMe, N1) /\ (T1, N3) := fTime(TMa, L0, N2) 
     /\ S := size(B0) /\ H0 := insert(((T0 plus (T1 plus (S * alpha)), N)), P) .
  
 crl [delay] :
     < L0 : agent | store : B0, set : SN > < L0 : process | UID : I0, command : (ask B1 -> C1) >
     < I : simulation | pqueue : T(Ra,((Ti, I0)), Le, Ri), flag : false, pend : P, Atts >
  => < L0 : agent | store : B0, set : SN > < L0 : process | UID : I0, command : (ask B1 -> C1) >
     < I : simulation | pqueue : merge(Le, Ri), flag : false, pend : H0, Atts >
  if not(entails(B0,B1)) /\ H0 := insert(((Ti, I0)),P) .
  
 crl [recursion] :
     < L0 : process | UID : I0, command : mu(N0, C0) >
     < I : simulation | pqueue : T(Ra,((Ti, I0)), Le, Ri), nextID : N, flag : false, pend : P, counter : N1, 
                        tTM : TMt, sTM : TMs, eTM : TMe, Atts >
  => < L0 : process | UID : N, command : replace(N0, C0, mu(N0,C0)) >
     < I : simulation | pqueue : T(Ra,((Ti, I0)), Le, Ri), nextID : (N + 1), flag : true, pend : H0, counter : N2, 
                      tTM : TMt, sTM : TMs, eTM : TMe, Atts >
  if (T0, N2) := getTimeCmd(C0, L0, TMt, TMs, TMe, N1) /\ H0 := insert(((T0, N)), P) .
  
 crl [search] :
     < L0 : agent | store : B0, set : SN >
     < L0 : process | UID : I0, command : watch(C0, B1) >
     < I : simulation | pqueue : T(Ra,((Ti, I0)), Le, Ri), nextID : N, flag : false, pend : P, counter : N1, 
                        tTM : TMt, sTM : TMs, eTM : TMe, Atts >
  => < L0 : agent | store : B0, set : SN >
     < L0 : process | UID : N, command : exc(LC, LF) >
     < I : simulation | pqueue : T(Ra,((Ti, I0)), Le, Ri), nextID : (N + 1), flag : true, pend : H0, counter : N'', 
                        tTM : TMt, sTM : TMs, eTM : TMe, Atts >
  if not(entails(B0,B1)) /\ LC := command-list(L0, SN, watch(C0, B1)) /\ P(LF, N') := prob-list(size(LC), N1) 
     /\ (T0, N'') := getTimeCmd(exc(LC, LF), L0, TMt, TMs, TMe, N') /\ H0 := insert(((T0, N)), P) .
  
 crl [found] :
     < L0 : agent | store : B0, set : SN >
     < L0 : process | UID : I0, command : watch(C0, B1) >
     < I : simulation | pqueue : T(Ra,((Ti, I0)), Le, Ri), nextID : N, flag : false, pend : P, counter : N1, 
                        tTM : TMt, sTM : TMs, eTM : TMe, Atts >
  => < L0 : agent | store : B0, set : SN >
     < L0 : process | UID : N, command : C0 >
     < I : simulation | pqueue : T(Ra,((Ti, I0)), Le, Ri), nextID : (N + 1), flag : true, pend : H0, counter : N2, 
                        tTM : TMt, sTM : TMs, eTM : TMe, Atts >
  if entails(B0, B1) /\ (T0, N2) := getTimeCmd(C0, L0, TMt, TMs, TMe, N1) /\ H0 := insert(((T0, N)), P) .
endm
\end{maude}
\subsection{apmaude.maude}
\label{app-apmaude}

\begin{maude}
mod APMAUDE is
  pr SCCP * (sort Configuration to Config) .
  pr COUNTER .
  
  var  Atts    : AttributeSet .
  vars H P     : Heap{2Tuple} .
  vars I N     : Nat .
  var  B       : Bool .
  var  L L0    : Aid .
  var  C B0    : Boolean .
  vars T T0    : Time .
  var  R       : Float .
  vars Conf X  : Config .
  var  P0      : Heap{2Tuple} .
  vars V1 V2   : AList{Aid} .
  var  Obj     : Object .
  
  ---- used by Quatex
  op flg : Bool Float -> Config . ---- a flag delimiting execution rounds
  
  op tick : Config -> Config .
  op tick2 : Config -> Config .
  eq tick(Conf) = tick2(Conf) .
  
  eq tick2( Conf flg(B, R) ) = Conf flg(true, R + 500.0) .
  
  op val : Nat Config -> Float .
  eq val(1, Conf) = getExcecutionTime(Conf) .
  
  ----- tick rule
 crl [tick] :
     < I : simulation | pqueue : P, gtime : T, flag : true, pend : P0, Atts >
     flg(true, R) Conf
  => < I : simulation | pqueue : merge(delta(deleteMin(P),T0),P0), gtime : (T plus T0), flag : false, 
                        pend : empty, Atts >
     flg(float(T plus T0) < R, R) Conf
  if T0 := p1(findMin(P))
     [print "tick " T] .
  
  ---- init state
  op initState : Nat -> Config .
  op initState : -> Config .
  rl initState => initState(counter) .
  eq initState(N) 
   = < root : agent | store : true >
     < root : process | UID : 1, command : ( ( ( exc( ( ( ( ( ind( ( ( tell(A:Integer === 1) ) in 1 ) 
       ( ( tell(B:Integer === 1) ) in 2 ) ( ( tell(C:Integer === 1) ) in 3 ) ( ( tell(D:Integer === 1) ) 
       in 4 ) , 0.5 0.5 0.5 0.5 ) ) in 4 ) || tell(Y:Integer === 5) || ask(Y:Integer > 2) -> 
       ( tell(Y:Integer > 2) out 1 ) ) in 1 ) ( ( tell(Y:Integer === 25) || ask(Y:Integer > 2) -> 
       ( tell(Y:Integer > 2) out 2 ) ) in 2 ), 0.60 0.40 ) || ask (Y:Integer > 2) -> 
       ( tell(X:Integer === 15) || ask(X:Integer >= 10) -> ( tell(X:Integer >= 10) out 1 ) ) ) in 1 ) ||
       ( ( ( ( tell(Z:Integer === 9) || ask(Z:Integer < 15) -> ( tell(Z:Integer < 15) out 3 ) ) in 3 ) || 
       ( ( tell(W:Integer === 25) || ask(W:Integer > 0) -> ( tell(W:Integer > 0) out 4 ) ) in 4 ) || 
       ask (Z:Integer < 15 and W:Integer > 0) -> ( tell(V:Integer === 67) || ask(V:Integer < 100) -> 
       ( tell(V:Integer < 100) out 2 ) ) ) in 2 ) ||  ( ask (X:Integer >= 10 and V:Integer < 100) -> 
       ( tell(U:Integer === 50) || ask(U:Integer < 55) -> tell(DONE:Boolean) ) ) ) >
     < 1 : simulation | gtime : 0,pqueue : T(1,((0,1)),empty,empty),pend : empty,nextID : 19, flag : false, 
                        counter : N, tTM : ((root) |-> Norm(1.0, 0.2)), aTM : ((root) |-> Norm(1.2, 0.2)) ,
                        sTM : ((root) |-> Norm(0.5, 0.2)) , eTM : ((root) |-> Norm(0.5, 0.2)), factor : 1/2 >
     flg(false, 0.0) .
  
  --- pvesta functions
  --- execution time
  op getExcecutionTime : Config -> Float .
  eq getExcecutionTime(< I : simulation | gtime : T, flag : false, pqueue : empty, Atts > Conf) = float(T) .
endm
\end{maude}

\subsection{formula.quatex}
\label{app-quatex}

\begin{maude}
  execTime ( ) = { s.rval( 1 ) } ; ;
  eval E[ # execTime ( ) ] ;
\end{maude}

\end{document}

%% file: packages.tex
\usepackage{amsmath}
\usepackage{amssymb}
\usepackage{amsthm}
\usepackage{enumerate}
\usepackage{enumitem}
\usepackage{fancyvrb}
\usepackage[T1]{fontenc}
\usepackage[utf8]{inputenc}
\usepackage{setspace}
\usepackage{tikz}
\usetikzlibrary{trees,arrows,arrows.meta,shapes,decorations.pathmorphing,backgrounds,positioning,fit,petri}
\usepackage{todonotes}
\usepackage[normalem]{ulem}   
\usepackage{url}
\usepackage[all]{xy}
\usepackage{bm}
\usepackage{mathtools}
\usepackage{caption}
\usepackage{subcaption}
\usepackage{relsize}
\usepackage{stmaryrd}
\usepackage{mathpartir}
\usepackage{graphicx}
\usepackage{float}
\usepackage{hyperref}
\usepackage{listings}
\usepackage{syntax}
\usepackage{geometry}
\usepackage{graphicx}
\usepackage{fleqn}
\usepackage{colortbl}

\graphicspath{{images/}}
\tikzset{%
	>={Latex[width=2mm,length=2mm]},
	module/.style = {rectangle, draw, minimum height=0.8cm, 
		minimum width=2.5cm, fill=orange!15, text centered, 
		font=\ttfamily},
	store/.style = {rectangle, draw, minimum height=1.4cm, 
		fill=orange!15, text centered, font=\ttfamily,
		minimum width=1.5cm},
  agent/.style = {circle, draw, minimum height=1cm, 
    fill=orange!15, text centered, font=\ttfamily,
    minimum width=1cm},
	state/.style = {circle, draw, minimum height=1.4cm, 
		fill=orange!15, text centered, font=\rmfamily,
		minimum width=1.5cm},
	triplearrow/.style={
		draw=black!75,
		color=black!75,
		double distance=3pt,
		postaction={draw=black!75, color=black!75}, 
		->},
	pil/.style={
		->,
		thick,
		shorten <=2pt,
		shorten >=2pt,},
}

\usepackage{proof} 
\usepackage{myproof}
\usepackage[inference]{semantic}

%% file: macros.tex
\newcommand{\QUATEX}{\textrm{QuaTEx}}
\newcommand{\PVESTA}{\textrm{PVeStA}}
\newcommand{\CS}{\textnormal{CS}}
\newcommand{\SCS}{\textnormal{SCS}}
\newcommand{\SCSE}{\textnormal{SCSE}}

\newcommand{\PTSCCP}{\textsf{sscc}}
\newcommand{\SSCC}{\textsf{sscc}}
\newcommand{\CCP}{\textnormal{CCP}}

\newcommand{\SCCP}{\textnormal{SCCP}}

\newcommand{\EQ}{=}
\newcommand{\IFS}{\mathbf{if}}
\newcommand{\REW}{\rightarrow}
\newcommand{\WPROB}{\textrm{with probability}}
\newcommand{\eq}[2]{#1 \EQ #2}

\newcommand{\crl}[3]{#1 \REW #2 \; \IFS\; #3}

\newcommand{\pcrl}[4]{#1 \REW #2 \; \IFS\; #3\;\WPROB\;#4}

\newcommand{\cond}{\gamma}

\newcommand{\tb}[1]{\textbf{#1}}

\newcommand{\ccrl}[3]{#1 \REW #2\;\textnormal{\bf if}\; {#3}}

\newcommand{\can}[2]{{#1}\!\downarrow_{#2}}

\newcommand{\ecal}{\mathcal{E}}
\newcommand{\lcal}{\mathcal{L}}
\newcommand{\rcal}{\mathcal{R}}
\newcommand{\tcal}{\mathcal{T}}

\newcommand{\ls}[1]{\mathit{ls}(#1)}
\newcommand{\ded}{\vdash}

\newcommand{\rews}{\rightarrow}
\newcommand{\func}[3]{#1 : #2 \longrightarrow #3}
\newcommand{\lang}{{0}}
\newcommand{\nlang}{{1}}
\newcommand{\oqff}[2]{\textit{\textit{QF}}_{#1}({#2})}

\DeclarePairedDelimiter{\efunc}{\uparrow}{}					

\newcommand{\true}{\it{true}}								
\newcommand{\false}{\it{false}}								
\newcommand{\Con}{\it{Con}}									
\newcommand{\join}[2]{#1 \ \sqcup \ #2}						
\newcommand{\cleq}[2]{#1 \sqsubseteq #2}					
\newcommand{\cgeq}[2]{#1 \sqsupseteq #2}					

\newcommand{\sfunc}[1]{\left[ #1 \right]}
\newcommand{\sfuncs}{\K{1}{\cdot},\dots,\K{n}{\cdot}}	
\newcommand{\efuncs}{\efunc{_1},\ldots,\efunc{_n}}				
\newcommand{\defsymbol}{\stackrel{\textup{\texttt{def}}}  {=}}	
\newcommand{\entails}{\vdash}

\newcommand{\Conf}[2]{\langle #1; #2\rangle}
\newcommand{\conf}[4]{\langle #1, #2, #3, #4 \rangle}

\newcommand{\extr}[2]{\uparrow_#1 (#2)}

\newcommand{\cde}[1]{\textnormal{\texttt{#1}}}
\newcommand{\rlname}[1]{\cde{[#1]}}

\lstdefinelanguage{Maude}{%
   keywords={
    , mod, fmod, endm, endfm
    , pr , protecting 
    , ex , extending 
    , inc, including
    , sort, sorts, subsort, subsorts
    , var, vars
    , op, ops
    , eq, ceq
    , rl, crl
    , if, then, else, fi
    , search
    , red, reduce
    }
}
\lstnewenvironment{maude}
{\lstset{language=Maude ,
    keywordstyle=\color{blue}
  , basicstyle=\ttfamily\singlespacing\scriptsize
  , commentstyle={}
  , columns=flexible
  , numbers=none
  , showstringspaces=false
  , keepspaces=true
  , aboveskip=-3pt
  } 
}
{}

\DefineVerbatimEnvironment{maude2}{Verbatim}{fontsize=\scriptsize}

\lstdefinelanguage{Sccp}{%
}

\lstnewenvironment{sccp}
{\lstset{language=SCCP
		, keywords={begin, end, var}	
		, keywordstyle=\color{red}
		, keywords=[2]{ask, tell, Int, Bool, x, v, r}
		, keywordstyle=[2]\color{blue}
		, basicstyle=\ttfamily\singlespacing\small
		, commentstyle={}
		, columns=flexible
		, numbers=none
		, showstringspaces=false
		, keepspaces=true
		, aboveskip=-3pt
		, breaklines=true
	} 
}
{}

\DefineVerbatimEnvironment{sccp2}{Verbatim}{fontsize=\scriptsize}

\newcommand{\tellp}[1]{\tell(#1)}
\newcommand{\askp}[2]{\ask \  #1 \  \rightarrow \ #2}

\newcommand{\parp}[2]{ #1 \, \parallel \, #2}

\newcommand{\skipp}{\Skip}

\renewcommand{\iff}{\mbox{\ \ iff \ \ }}

\newcommand{\ask}{{\bf ask}}

\newcommand{\tell}{{\bf tell}}

\newcommand{\Stop}{{\bf 0}}
\newcommand{\Skip}{{\bf skip}}

\newcommand{\pairccp}[2]{\Conf{#1}{#2}}

\long\def\comment#1{}

\newcommand{\To}{\Rightarrow}

\newcommand{\K}[2]{\left[ #2 \right] _{#1}}			

\newcommand{\prj}[2]{#1^{{#2}}}

\makeatletter

\newdimen\w@dth

\def\setw@dth#1#2{\setbox\z@\hbox{\scriptsize $#1$}\w@dth=\wd\z@
\setbox\@ne\hbox{\scriptsize $#2$}\ifnum\w@dth<\wd\@ne \w@dth=\wd\@ne \fi
\advance\w@dth by 1.2em}

\def\t@^#1_#2{\allowbreak\def\n@one{#1}\def\n@two{#2}\mathrel
{\setw@dth{#1}{#2}
\mathop{\hbox to \w@dth{\rightarrowfill}}\limits
\ifx\n@one\empty\else ^{\box\z@}\fi
\ifx\n@two\empty\else _{\box\@ne}\fi}}
\def\t@@^#1{\@ifnextchar_ {\t@^{#1}}{\t@^{#1}_{}}}

\def\t@left^#1_#2{\def\n@one{#1}\def\n@two{#2}\mathrel{\setw@dth{#1}{#2}
\mathop{\hbox to \w@dth{\leftarrowfill}}\limits
\ifx\n@one\empty\else ^{\box\z@}\fi
\ifx\n@two\empty\else _{\box\@ne}\fi}}
\def\t@@left^#1{\@ifnextchar_ {\t@left^{#1}}{\t@left^{#1}_{}}}

\def\two@^#1_#2{\def\n@one{#1}\def\n@two{#2}\mathrel{\setw@dth{#1}{#2}
\mathop{\vcenter{\hbox to \w@dth{\rightarrowfill}\kern-1.7ex
                 \hbox to \w@dth{\rightarrowfill}}%
       }\limits
\ifx\n@one\empty\else ^{\box\z@}\fi
\ifx\n@two\empty\else _{\box\@ne}\fi}}
\def\tw@@^#1{\@ifnextchar_ {\two@^{#1}}{\two@^{#1}_{}}}

\def\tofr@^#1_#2{\def\n@one{#1}\def\n@two{#2}\mathrel{\setw@dth{#1}{#2}
\mathop{\vcenter{\hbox to \w@dth{\rightarrowfill}\kern-1.7ex
                 \hbox to \w@dth{\leftarrowfill}}%
       }\limits
\ifx\n@one\empty\else ^{\box\z@}\fi
\ifx\n@two\empty\else _{\box\@ne}\fi}}
\def\t@fr@^#1{\@ifnextchar_ {\tofr@^{#1}}{\tofr@^{#1}_{}}}

\newdimen\W@dth
\def\setW@dth#1#2{\setbox\z@\hbox{$#1$}\W@dth=\wd\z@
\setbox\@ne\hbox{$#2$}\ifnum\W@dth<\wd\@ne \W@dth=\wd\@ne \fi
\advance\W@dth by 1.2em}

\def\T@^#1_#2{\allowbreak\def\N@one{#1}\def\N@two{#2}\mathrel
{\setW@dth{#1}{#2}
\mathop{\hbox to \W@dth{\rightarrowfill}}\limits
\ifx\N@one\empty\else ^{\box\z@}\fi
\ifx\N@two\empty\else _{\box\@ne}\fi}}
\def\T@@^#1{\@ifnextchar_ {\T@^{#1}}{\T@^{#1}_{}}}

\def\T@left^#1_#2{\def\N@one{#1}\def\N@two{#2}\mathrel{\setW@dth{#1}{#2}
\mathop{\hbox to \W@dth{\leftarrowfill}}\limits
\ifx\N@one\empty\else ^{\box\z@}\fi
\ifx\N@two\empty\else _{\box\@ne}\fi}}
\def\T@@left^#1{\@ifnextchar_ {\T@left^{#1}}{\T@left^{#1}_{}}}

\def\Tofr@^#1_#2{\def\N@one{#1}\def\N@two{#2}\mathrel{\setW@dth{#1}{#2}
\mathop{\vcenter{\hbox to \W@dth{\rightarrowfill}\kern-1.7ex
                 \hbox to \W@dth{\leftarrowfill}}%
       }\limits
\ifx\N@one\empty\else ^{\box\z@}\fi
\ifx\N@two\empty\else _{\box\@ne}\fi}}
\def\T@fr@^#1{\@ifnextchar_ {\Tofr@^{#1}}{\Tofr@^{#1}_{}}}

\def\Two@^#1_#2{\def\N@one{#1}\def\N@two{#2}\mathrel{\setW@dth{#1}{#2}
\mathop{\vcenter{\hbox to \W@dth{\rightarrowfill}\kern-1.7ex
                 \hbox to \W@dth{\rightarrowfill}}%
       }\limits
\ifx\N@one\empty\else ^{\box\z@}\fi
\ifx\N@two\empty\else _{\box\@ne}\fi}}
\def\Tw@@^#1{\@ifnextchar_ {\Two@^{#1}}{\Two@^{#1}_{}}}

\def\to{\@ifnextchar^ {\t@@}{\t@@^{}}}
\def\from{\@ifnextchar^ {\t@@left}{\t@@left^{}}}
\def\two{\@ifnextchar^ {\tw@@}{\tw@@^{}}}
\def\tofro{\@ifnextchar^ {\t@fr@}{\t@fr@^{}}}
\def\To{\@ifnextchar^ {\T@@}{\T@@^{}}}
\def\From{\@ifnextchar^ {\T@@left}{\T@@left^{}}}
\def\Two{\@ifnextchar^ {\Tw@@}{\Tw@@^{}}}
\def\Tofro{\@ifnextchar^ {\T@fr@}{\T@fr@^{}}}

\makeatother

\newtheorem{example}{Example}
\newtheorem{definition}{Definition}

\newcommand{\delayp}[2]{\eta(#1,#2)}
\newcommand{\delays}[2]{\eta_s(#1,#2)}